\documentclass[12pt]{article}
\newcommand{\filename}{2LM-M\_quadratic\_descriptors\_v10}
\usepackage{authblk}
\usepackage{amsfonts}
\usepackage{amssymb}
\usepackage{amsmath}
\usepackage{amsthm}
\usepackage{framed}

\usepackage{enumitem}
\usepackage{geometry}
\usepackage{url}
\usepackage{booktabs}

\usepackage{multirow}

\usepackage{mdframed}
\usepackage{mathtools}
\usepackage{algorithm}  
\usepackage{algorithmic}

\newcommand{\Ctrain}{\mathcal{C}_\mathrm{train}}  
\newcommand{\Ctest}{\mathcal{C}_\mathrm{test}}

\newcommand{\lf}{\mathrm{lf}}

\newcommand{\eledeg}{\mathrm{eledeg}}

\newcommand{\ttH}{{\tt H}}  
\newcommand{\ttC}{{\tt C}}  
\newcommand{\ttO}{{\tt O}}  
\newcommand{\ttN}{{\tt N}}  
\newcommand{\ttP}{{\tt P}}  
\newcommand{\ttF}{{\tt F}}  
\newcommand{\ttCl}{{\tt Cl}}  
\newcommand{\ttS}{{\tt S}}  
\newcommand{\ttSi}{{\tt Si}}  
\newcommand{\ttPb}{{\tt Pb}}  
\newcommand{\ttBr}{{\tt Br}}  
\newcommand{\ttI}{{\tt I}}

\newcommand{\oH}{\overline{{\tt H}}}

\newcommand{\C}{\mathbb{C}}  
\newcommand{\Co}{\mathbb{C}}

\newcommand{\anC}{\langle \mathbb{C} \rangle}  
\newcommand{\anpsi}{\langle \psi \rangle}  

\newcommand{\VH}{V_{\tt H}}

\newcommand{\R}{\mathbb{R}} 
 
\newcommand{\RK}{\mathbb{R}^K}

\newcommand{\dcp}{\mathrm{dcp}}

\newcommand{\Vleaf}{V_\mathrm{leaf}} 
\newcommand{\Eleaf}{E_\mathrm{leaf}} 

\newcommand{\sint}{\sigma_\mathrm{int}} 
\newcommand{\sce}{\sigma_\mathrm{ce}}

\newcommand{\Ez}{E_{(0/1)}}
\newcommand{\Ew}{E_{(\geq 1)}}
\newcommand{\Et}{E_{(\geq 2)}}
\newcommand{\Eew}{E_{(=1)}}

\newcommand{\Gac}{\Gamma_\mathrm{ac}}

\newcommand{\w}{w}
\newcommand{\x}{x}

\newcommand{\ta}{{\tt a}}
\newcommand{\tb}{{\tt b}}
\newcommand{\Ldg}{\Lambda_{\mathrm{dg}}}

\newcommand{\fc}{\mathrm{fc}}

\newcommand{\val}{\mathrm{val}}

\newcommand{\inte}{\mathrm{int}}

\newcommand{\nint}{\mathrm{n}^\mathrm{int}}

\newcommand{\h}{\mathrm{ht}}

\newcommand{\cs}{\mathrm{cs}}
\newcommand{\ch}{\mathrm{ch}}

\newcommand{\dg}{\mathrm{dg}}

\newcommand{\na}{\mathrm{na}}

\newcommand{\acC}{\mathrm{ac}_\mathrm{C}}

\newcommand{\ns}{\mathrm{ns}}

\newcommand{\ec}{\mathrm{ec}}
\newcommand{\ac}{\mathrm{ac}}

\newcommand{\bl}{\mathrm{bl}}

\newcommand{\bd}{\mathrm{bd}}

\newcommand{\UB}{\mathrm{UB}}
\newcommand{\LB}{\mathrm{LB}}

\newcommand{\ex}{\mathrm{ex}}

\newcommand{\GC}{G_\mathrm{C}}

\newcommand{\VC}{V_\mathrm{C}}

\newcommand{\EC}{E_\mathrm{C}}

\begin{document} 

\begin{center}
   {\Large\bf 
    Molecular Design Based on 
    Integer Programming and Quadratic Descriptors
    in a Two-layered Model}
   \\ 
\end{center}

\begin{center} 
Jianshen Zhu$^1$, 
Naveed Ahmed Azam$^1$, 
Shengjuan Cao$^1$,  
Ryota Ido$^1$,  
Kazuya Haraguchi$^{1}$, 
Liang Zhao$^2$, 
Hiroshi Nagamochi$^1$ 
 and  
 Tatsuya Akutsu$^3$ 
\end{center} 
%
%
{\small 
 $^1$  Department of Applied Mathematics and Physics, Kyoto University, 
 Kyoto 606-8501, Japan\\
$^2$   Graduate School of Advanced Integrated Studies in Human Survavibility
     (Shishu-Kan),   Kyoto University, Kyoto 606-8306, Japan \\
$^3$   Bioinformatics Center,  Institute for Chemical Research, 
  Kyoto University, Uji 611-0011, Japan 
}

\begin{quote}  
{\bf Abstract}\\  
A novel framework has recently been proposed for designing 
the molecular structure of chemical compounds
with a desired chemical property,
where design of novel drugs is an important topic in bioinformatics
and chemo-informatics.
The framework infers a desired chemical graph
by  solving  a mixed integer linear program (MILP)
that simulates the computation process of
a feature function  defined by a two-layered model on chemical graphs
and a prediction function constructed by a machine learning method. 
A set of graph theoretical descriptors in the feature function
plays a key role to derive a compact formulation of such an MILP. 
To improve the learning performance of prediction functions in the framework
maintaining the compactness of the MILP,
 this paper utilizes the product of two of those descriptors 
as a new descriptor
and then  designs  a method of reducing the number of descriptors.
The results of our computational experiments suggest that  
 the proposed method improved the learning performance 
 for many chemical properties  
 and can infer a chemical structure 
with  up to 50 non-hydrogen atoms.



\end{quote}

\section{Introduction}\label{sec:introduction}
 
\noindent {\bf Background~}
In recent years, extensive studies have been done on 
design of novel molecules using various machine learning techniques
\cite{Lo18,Tetko20}.
Computational molecular design has also a long history in 
the field of chemo-informatics, and has been studied under the names
of \emph{quantitative structure activity relationship} (QSAR) \cite{CMF14}
and
\emph{inverse quantitative structure activity relationship}
(inverse QSAR) \cite{Miyao16,Ikebata17,Rupakheti15}.
This design problem has also become a hot topic in both bioinformatics and
machine learning.

The purpose of QSAR is to predict chemical activities from
given chemical structures \cite{CMF14}.
In most of QSAR studies, a chemical structure is represented as
a vector of real numbers called \emph{features} or \emph{descriptors}
and then a prediction function is applied to the vector,
where a chemical structure is given as 
an undirected graph called a \emph{chemical graph}.
A prediction function is usually obtained 
from existing structure-activity relation data.
To this end,
various regression-based methods have been utilized in traditional QSAR studies,
whereas machine learning-based methods, including
artificial neural network (ANN)-based methods, have recently been utilized
\cite{Ghasemi18,kim22}.

Conversely,
the purpose of inverse QSAR is to predict chemical structures
from given chemical activities \cite{Miyao16,Ikebata17,Rupakheti15},
where additional constraints may often be imposed to effectively restrict
the possible structures.
In traditional inverse QSAR,
a feature vector is firstly computed by applying some optimization
or sampling method to the prediction function obtained by usual QSAR
and then chemical structures are reconstructed from the feature vector.
However, the reconstruction itself is quite difficult 
because the number of possible chemical graphs is huge~\cite{BMG96}. 
Indeed, it is NP-hard to infer a chemical graph from a given feature vector
except for some simple cases~\cite{AFJS12}.  
Due to this inherent difficulty,
most existing methods employ heuristic methods for reconstruction
of chemical structures and thus
do not guarantee optimal or exact solutions.

On the other hand, 
one of the advantages of ANNs is that generative models are available,
such as autoencoders and generative adversarial networks.
Furthermore, graph structured data can be directly handled by
using graph convolutional networks~\cite{Kipf16}.
Therefore, it is reasonable to try to apply ANNs to inverse QSAR
\cite{xiong22}.
Indeed, various ANN  models have been applied, which includes
recurrent neural networks~\cite{Segler18,Yang17}, 
variational autoencoders~\cite{Gomez18}, 
grammar variational autoencoders~\cite{Kusner17},
generative adversarial networks~\cite{DeCao18},
and invertible flow models~\cite{Madhawa19,Shi20}.
However,
optimality or exactness of the solutions is not yet guaranteed by
these methods.

\begin{figure}[!ht]  \begin{center}
\includegraphics[width=.77\columnwidth]{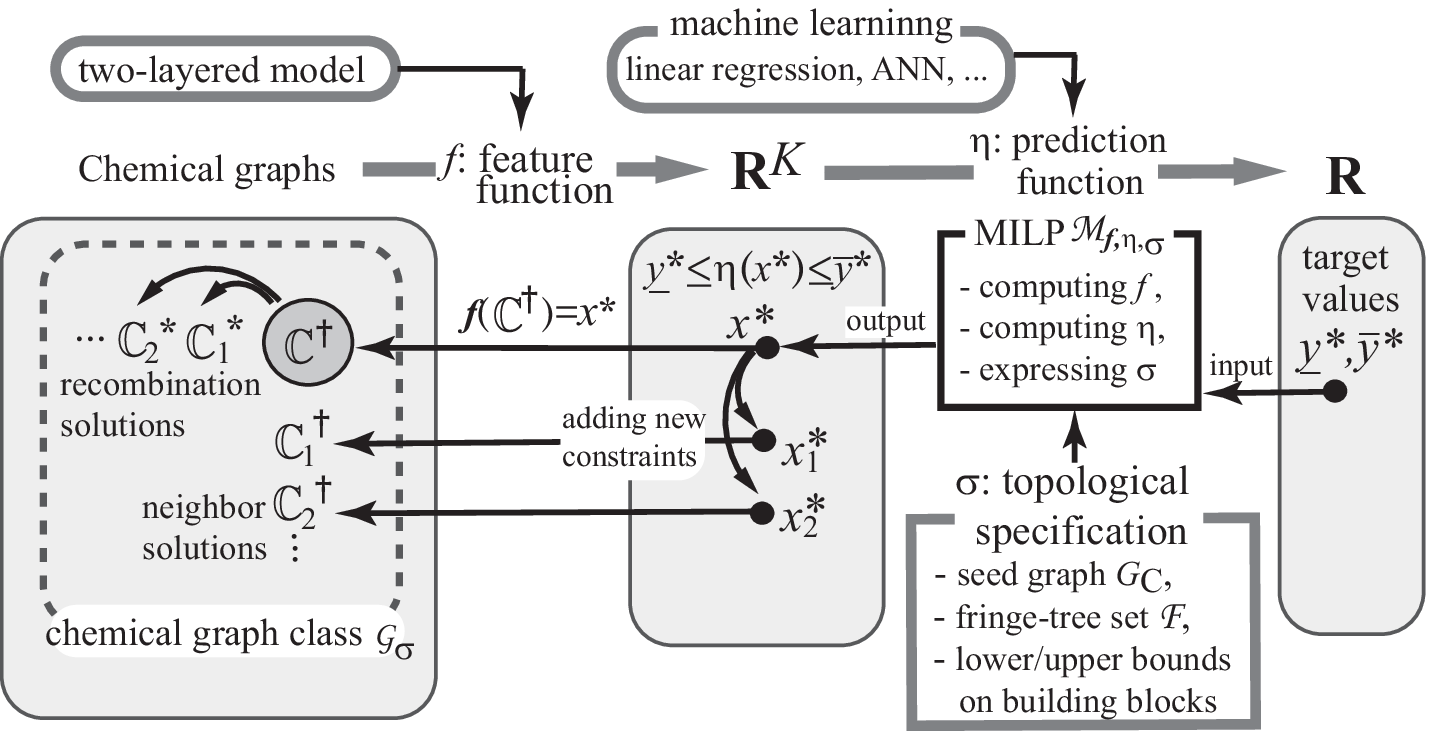}
\end{center} \caption{An illustration of inferring
desired chemical graphs $\Co\in \mathcal{G}_\sigma$ with
$\underline{y^*}\leq \eta(f(\Co))\leq \overline{y^*}$.   } 
\label{fig:framework}  \end{figure}    

\smallskip
\noindent {\bf Framework~}
A novel  framework for inferring chemical graphs has been developed
 \cite{SZAHZNA21,ZAHZNA21,poly_ICZAHZNA22,gridAZHZNA21} based on an idea of formulating
 as a mixed integer linear programming (MILP)
 the computation process of a prediction function constructed
by a machine learning method.
It consists of two main phases: the first phase 
 constructs a prediction function $\eta$ for a chemical property and the second phase 
infers a chemical graph with a target value of the property based on the function $\eta$. 
For a chemical property $\pi$, let $\mathcal{C}_{\pi}$ be a data set  of chemical graphs  
such that the observed value $a(\Co)$ of property $\pi$ 
 for every  chemical graph $\Co \in \mathcal{C}_{\pi}$ is available.
 In the first phase, we introduce a feature function $f: \mathcal{G}\to \mathbb{R}^K$ 
for a positive integer $K$,
where the descriptors of
a chemical graph are defined based on local graph  structures in a special way
called a {\em two-layered model}. 
We then construct a prediction function $\eta$
by a machine learning method such as linear regression, decision tree and an ANN
so that the output $y=\eta(x)\in \mathbb{R}$ of    
the feature vector  $x=f(\Co)\in \mathbb{R}^K$ for each $\Co\in \mathcal{C}_{\pi}$
  serves as a predicted value to the real value $a(\Co)$.  
In the second phase of inferring a desired chemical graph,
we specify not only a target chemical value for property $\pi$ but also
an abstract structure for a chemical graph to be inferred.
The latter is described by a set of rules based on the two-layered model called 
a {\em topological specification} $\sigma$, and denote by $\mathcal{G}_\sigma$
 the set of all chemical graphs that satisfy the rules in $\sigma$. 
The users select topological specification  $\sigma$ and
 two reals $\underline{y}^*$ and $\overline{y}^*$ 
 as an interval  for a  target chemical value.
The task of the  second phase is 
to infer  chemical graphs $\Co^*\in \mathcal{G}_\sigma$
such that  
 $\underline{y}^*\leq \eta(f(\C^*))\leq \overline{y}^*$
(see Figure~\ref{fig:framework}  for an illustration).
For this, we formulate an MILP $\mathcal{M}_{f,\eta,\sigma}$
that represents 
(i) the  computation process of  $x:=f(\Co)$ from a chemical graph $\Co$
in the feature function $f$;
(ii) that of  $y:=\eta(x)$ from a vector $x\in\mathbb{R}^K$
in the prediction function $\eta$; and
(iii) the  constraint  $\Co\in  \mathcal{G}_\sigma$.
Given an interval with $\underline{y}^*,\overline{y}^* \in \mathbb{R}$,
 we solve the MILP $\mathcal{M}_{f,\eta,\sigma}$
to find  a feature vector $x^*\in \mathbb{R}^K$
 and a chemical graph $\Co^{\dagger}\in  \mathcal{G}_\sigma$ 
 such that $f(\Co^\dagger)=x^*$ and  
$\underline{y}^*\leq \eta(x^*) \leq \overline{y}^*$
(where if the MILP instance is infeasible
 then this suggests that $\mathcal{G}_\sigma$ 
does  not contain such a desired chemical graph). 
In the second phase, we next generate 
  some other desired chemical graphs based on the solution $\Co^{\dagger}$.
For this, the following two methods have been designed. 

 The first method  constructs 
   isomers of   $\C^\dagger$   without solving any new MILP.
In this method, we first decompose  the chemical graph $\C^\dagger$ into 
a set of chemical acyclic graphs $T^\dagger_1,T^\dagger_2,\ldots,T^\dagger_q$,
and next construct  a set $\mathcal{T}_i$ of isomers $T^*_i$ of each tree $T^\dagger_i$
 such that $f(T^*_i)=f(T^\dagger_i)$ by a dynamic programming algorithm
 due to Azam~et~al.~\cite{AZSSSZNA20}.  
 Finally 
 we choose an isomer $T^*_i\in \mathcal{T}_i$ for each $i=1,2,\ldots,q$
 and  assemble them into an isomer $\C^*\in \mathcal{G}_\sigma$ of $\C^\dagger$
 such that $f(\C^*)=x^*=f(\C^\dagger)$.
  The first method generates such isomers $\C_1^*, \C_2^*, \ldots$ 
 which
 we call {\em  recombination solutions} of $\C^\dagger$.

The second method constructs   new solutions 
by solving the MILP $\mathcal{M}_{f,\eta,\sigma}$
with an additional set $\Theta$ of new linear constraints~\cite{gridAZHZNA21}.
We first prepare arbitrary $p_{\mathrm{dim}}$
 linear functions $\theta_j: \RK\to \R, j=1,2,\ldots,p_{\mathrm{dim}}$
and consider a neighbor of $\C^\dagger$ 
 defined by a set of chemical graphs $\C^*$ that satisfy linear constraints 
 $k\delta\leq |\theta_j(f(\C^*))-\theta_j(f(\C^\dagger))|\leq (k+1)\delta,  
 j=1,2,\ldots,p_{\mathrm{dim}}$ 
 for  a small real  $\delta>0$ and an integer $k\geq 1$.  
By changing the  integer $k$ systematically, we can search for new solutions
$\C^\dagger_1,\C^\dagger_2,  \ldots \in \mathcal{G}_\sigma$ 
of  MILP $\mathcal{M}_{f,\eta,\sigma}$
with constraint $\Theta$ such that the feature
vectors $x^*=f(\C^\dagger), x^*_1=f(\C^\dagger_1), x^*_2=f(\C^\dagger_2),\ldots$ 
are all slightly different.
We call these chemical graphs
 $\C^\dagger_1,\C^\dagger_2,  \ldots$ {\em neighbor solutions} of $\C^\dagger$,
 where a neighbor solution is not an isomer of $\C^\dagger$. 


The main reason why the framework can infer a chemical compound
with 50 non-hydrogen atoms is that  the descriptors of
a chemical graph are defined  on local graph structures in the two-layered model  and
thereby an MILP necessary to represent
a chemical graph can be formulated 
as a considerably compact form that is efficiently solvable by a standard solver.

\smallskip
\noindent {\bf Contribution~}
In the framework, all descriptors $x(1),x(2),\ldots,x(K)$ in the feature vector $x=f(\Co)$ 
are mainly the frequencies of local graph structures
based on the two-layered model
by which a chemical graph $\Co$ is regarded as a pair
of interior and exterior structures (see Section~\ref{sec:2LM} for details).
To derive a compact MILP formulation to infer a chemical graph,
it is important to use the current definition of descriptors.
However, there are some chemical properties for which
the performance of a prediction function constructed 
with the feature function $f$ remains rather low.
To improve the learning performance with the same two-layered model,
we add as a new descriptor the product $x(i)x(j)$ (or $x(i)(1-x(j))$ of two descriptors
(where each descriptor is assumed to be normalized within 0 and 1) and 
  call such a new descriptor a {\em quadratic descriptor}. 
This drastically increases the number of descriptors, which would take
extra running time in learning or cause overfitting to the data set.
Moreover,  computing quadratic descriptors  cannot be directly formulated as 
a set of linear constraints in the original MILP.
For this, we introduce a method of reducing a set of descriptors
into a smaller set that delivers a prediction function with a higher performance.
We also design an MILP formulation for representing a quadratic term   $x(i)x(j)$.
Based on the same MILP  $\mathcal{M}_{f,\eta,\sigma}$ formulation 
proposed by  Zhu~et~al.~\cite{ZAHZNA21},
we implemented the framework to treat  the feature function
with quadratic descriptors.
From the results of our computational experiments on over 40 chemical properties,
we observe that our new method of utilizing quadratic descriptors
improved the performance of a prediction function for many chemical properties.  

The paper is organized as follows.  
Section~\ref{sec:preliminary} introduces some notions on graphs and 
 a modeling of chemical compounds. 
Section~\ref{sec:2LM} reviews  the two-layered model and
 a basic idea of descriptors by the model.
Section~\ref{sec:compute_Q} introduces a formulation for 
  computing a quadratic descriptor  in an MILP. 
Section~\ref{sec:experiment} reports the results on  computational 
experiments conducted for 42 chemical properties 
such as critical pressure, dissociation constants and  lipophilicity for monomers
and characteristic ratio and  refractive index for polymers. 
Section~\ref{sec:conclude} makes some concluding remarks.   
Some technical details are given in Appendices:   
 Appendix~\ref{sec:descriptor}
    for  all descriptors in our feature function; 
Appendix~\ref{sec:reduce_descriptors} for how to reduce the number of descriptors;
 Appendix~\ref{sec:specification}
   for a full description of a topological specification; and 
Appendix~\ref{sec:test_instances}
  for the detail of test instances used in our computational experiment.
  
%

\section{Preliminary}\label{sec:preliminary}

This section  introduces some notions and terminologies on graphs,
  modeling of chemical compounds and our choice of descriptors. 
 
Let $\mathbb{R}$, $\mathbb{R}_+$, $\mathbb{Z}$  and $\mathbb{Z}_+$ 
denote the sets of reals,  non-negative reals, 
integers and non-negative integers, respectively.
For two integers $a$ and $b$, let $[a,b]$ denote the set of 
integers $i$ with $a\leq i\leq b$.
For a vector $x\in \R^p$, the $j$-th entry of $x$ is denoted by $x(j)$.

\bigskip\noindent
{\bf  Graph} 
Given a  graph $G$, let $V(G)$ and $E(G)$ denote the sets
of vertices and edges, respectively.     
For a subset $V'\subseteq V(G)$ (resp., $E'\subseteq E(G))$ of
a graph $G$, 
let $G-V'$ (resp., $G-E'$) denote the graph obtained from $G$
by removing the vertices in $V'$ (resp.,  the edges in $E'$),
where we remove all edges incident to a vertex in $V'$ to obtain $G-V'$. 
%
A path with two end-vertices $u$ and $v$ is called a {\em $u,v$-path}. 
 
 We define a {\em rooted} graph to be
 a graph with a  designated vertex, called a {\em root}. 
 %
%
 For a graph $G$ possibly with a root,  
 a {\em leaf-vertex} is defined to be a non-root vertex 
 with degree 1.
 Call  the edge $uv$ incident to a leaf vertex $v$ a {\em leaf-edge},
 and denote by $\Vleaf(G)$ and $\Eleaf(G)$
  the sets of leaf-vertices and leaf-edges  in $G$, respectively.
 For a graph  or a rooted graph $G$,
 we define graphs $G_i, i\in \mathbb{Z}_+$ obtained from $G$
 by removing the set of leaf-vertices $i$ times so that
\[ G_0:=G; ~~ G_{i+1}:=G_i - \Vleaf(G_i), \]
where we call a vertex $v$ a {\em tree vertex} if $v\in \Vleaf(G_i)$
for some $i\geq 0$. 
Define the {\em height} $\h(v)$ of each tree vertex $v\in \Vleaf(G_i)$
to be $i$; and 
$\h(v)$ of each non-tree vertex $v$ adjacent to a tree vertex 
to be $\h(u)+1$ for the maximum $\h(u)$ of a tree vertex $u$ adjacent to $v$,
where we do not define height of any non-tree vertex not adjacent to any tree vertex. 
We call a vertex $v$ with $\h(v)=k$ a {\em leaf $k$-branch}.
The {\em height} $\h(T)$ of a rooted tree $T$ is defined
to be the maximum of $\h(v)$ of a vertex $v\in V(T)$. 
 
\subsection{Modeling of Chemical Compounds}\label{sec:chemical_model}

We review a modeling of chemical compounds introduced 
by  Zhu~et~al.~\cite{ZAHZNA21}. 

To represent a chemical compound, 
we introduce a set  of   chemical elements such as 
  {\tt H} (hydrogen),  {\tt C} (carbon), {\tt O} (oxygen), {\tt N} (nitrogen)  and so on.
 To distinguish a chemical element $\ta$ with multiple valences such as {\tt S} (sulfur),
 we denote a chemical element $\ta$ with a valence $i$ by $\ta_{(i)}$,
 where we do not use such a suffix $(i)$ 
 for a chemical element $\ta$ with a unique valence. 
Let $\Lambda$ be a set of chemical elements $\ta_{(i)}$.
For example,  $\Lambda=\{\ttH,  \ttC, \ttO, \ttN, \ttP, \ttS_{(2)}, \ttS_{(4)}, \ttS_{(6)}\}$. 
Let $\val: \Lambda\to [1,6]$ be a valence function.
For example, $\val(\ttH)=1$, $\val(\ttC)=4$, $\val(\ttO)=2$, $\val(\ttP)=5$,
$\val(\ttS_{(2)})=2$, $\val(\ttS_{(4)})=4$ and $\val(\ttS_{(6)})=6$.
 For each  chemical element $\ta\in \Lambda$, 
let $\mathrm{mass}(\ta)$  denote the mass   of  $\ta$.

A chemical compound  is represented by a {\em chemical graph} defined to be
a tuple $\Co=(H,\alpha,\beta)$  of
  a simple, connected undirected graph $H$ and  
    functions   $\alpha:V(H)\to \Lambda$  and  $\beta: E(H)\to [1,3]$.
The set of atoms and the set of bonds in the compound 
are represented by the vertex set $V(H)$ and the edge set $E(H)$, respectively.
The chemical element assigned to a vertex $v\in V(H)$
is represented by $\alpha(v)$ and 
 the bond-multiplicity  between two adjacent vertices  $u,v\in V(H)$
is represented by $\beta(e)$ of the edge $e=uv\in E(H)$.
We say that two tuples $(H_i,\alpha_i,\beta_i), i=1,2$ are
{\em isomorphic} if they admit an isomorphism $\phi$,
i.e.,  a bijection $\phi: V(H_1)\to V(H_2)$
such that
 $uv\in E(H_1), \alpha_1(u)=\ta, \alpha_1(v)=\tb, \beta_1(uv)=m$
 $\leftrightarrow$  
 $\phi(u)\phi(v) \in E(H_2), \alpha_2(\phi(u))=\ta, 
 \alpha_2(\phi(v))=\tb, \beta_2(\phi(u)\phi(v))=m$. 
 When $H_i$ is rooted at a vertex $r_i,  i=1,2$,
these chemical graphs $(H_i,\alpha_i,\beta_i),  i=1,2$ are
{\em rooted-isomorphic} (r-isomorphic) if 
they admit  an isomorphism $\phi$ such that $\phi(r_1)=r_2$. 

 For a notational convenience, we  use
 a function $\beta_\Co: V(H)\to [0,12]$ 
 for a chemical graph $\Co=(H,\alpha,\beta)$
  such that $\beta_\Co(u)$ means the sum of bond-multiplicities
 of edges incident to a vertex $u$; i.e., 
\[ \beta_\Co(u) \triangleq \sum_{uv\in E(H) }\beta(uv) 
\mbox{ for each vertex $u\in V(H)$.}\]
For each vertex $u\in V(H)$, 
 define the {\em electron-degree} $\eledeg_\Co(u)$  to be 
\[  \eledeg_\Co(u) \triangleq  \beta_\Co(u) - \val(\alpha(u)). \]
For each  vertex $u\in V(H)$, let $\deg_\Co(v)$ denote 
the number of vertices adjacent to $u$ in $\Co$. 
  
  For a chemical   graph  $\Co=(H,\alpha,\beta)$, 
  let  $V_{\ta}(\Co)$, $\ta\in \Lambda$
   denote the set of vertices $v\in V(H)$ such that $\alpha(v)=\ta$ in $\Co$
  and define the {\em hydrogen-suppressed chemical graph} $\anC$ 
to be  the graph obtained from $H$ by
  removing all the vertices $v\in \VH(\Co)$.

\subsection{Prediction Functions}\label{sec:prediction_functions}

Let $\mathcal{C}$ be a data set   of chemical graphs $\C$ with
an observed value $a(\C)\in \R$. 
Let $D$ be a set of descriptors and 
 $f$ be a feature function that maps a chemical graph $\C$
to a vector $ f(\C)\in \R^D$,
where $x(d)$ denotes the value of descriptor $d\in D$.
For a notational simplicity,
  we denote  $a_i=a(\C_i)$ and  $\x_i= f(\C_i)$ 
for an indexed graph $\C_i\in \mathcal{C}$. 

\subsubsection{Evaluation}\label{sec:evaluation}  

For  a prediction function $\eta: \R^D\to \R$, 
define an error function 
\[ \mathrm{Err}(\eta;\mathcal{C})  \triangleq 
\sum_{\C_i\in \mathcal{C}}(a_i - \eta(f(\C_i)))^2
=\sum_{\C_i\in \mathcal{C}}(a_i - \eta(\x_i))^2, \]
and define the {\em coefficient of determination}
 $\mathrm{R}^2(\eta,\mathcal{C})$ 
  to be 
\[ \displaystyle{ \mathrm{R}^2(\eta,\mathcal{C})\triangleq 
  1- \frac{\mathrm{Err}(\eta;\mathcal{C}) } 
  {\sum_{ \C_i\in \mathcal{C}  } (a_i-\widetilde{a})^2}  
  \mbox{   for  }
   \widetilde{a}= \frac{1}{|\mathcal{C}|}\sum_{ \C\in \mathcal{C} }a(\C).  } \] 
   
We evaluate a method of constructing a prediction function
over a set $D$ of descriptors by 5-fold cross-validation as follows.   
A single run $r$ of 5-fold cross-validation executes the following: 
 Partition a data set $\mathcal{C}$  randomly 
 into five subsets $\mathcal{C}^{(k)}$, $k\in[1,5]$ so that 
 the difference between  $|\mathcal{C}^{(i)}|$ and $|\mathcal{C}^{(j)}|$ is at most 1. 
 For each $k\in[1,5]$, let $\Ctrain:=\mathcal{C} \setminus \mathcal{C}^{(k)}$,
  $\Ctest:=\mathcal{C}^{(k)}$ and
 execute the method to construct a prediction function
  $\eta^{(k)}: \R^D\to \R$ over  
 a training set  $\Ctrain$
 and compute $g_r^{(k)}:=\mathrm{R}^2(\eta^{(k)},\Ctest)$.
Let $\mathrm{R}^2_{\mathrm{CV}}(\mathcal{C},D,p)$ denote 
 the median  of $\{g_{r_i}^{(k)}\mid k\in[1,5],i\in[1,p]\}$ of  
 $p$ runs $r_1,r_2,\ldots,r_p$ of 5-fold cross-validation.

\subsubsection{Linear regressions}\label{sec:linear_regression}  

For  a set $D$ of descriptors, a hyperplane is defined to be 
a pair  $(\w,b)$ of a vector $\w\in \R^D$ and a real $b\in \R$.
Given a hyperplane $(\w,b)$,
a prediction function $\eta_{\w,b}:\R^D\to \R$ is defined by setting
\[ \eta_{\w,b}(\x) \triangleq \w\cdot \x +b=\sum_{d\in D}\w(d)\x(d) +b. \]


Given a data set $\mathcal{C}$ and a set $D$ of descriptors, 
{\em multidimensional linear regression} MLR$(\mathcal{C},D)$ returns
a hyperplane $(\w,b)$ with $\w\in \R^D$  
 that minimizes $\mathrm{Err}(\eta_{\w,b}; \mathcal{C})$.
However, such a hyperplane $(\w,b)$ may contain
unnecessarily many non-zero reals $\w(d)$.
To avoid this, a minimization with an additional penalty term $\tau$ 
to the error function has been proposed.
Among them,     
 a Lasso function~\cite{Lasso96} is defined to be 
\[\frac{1}{2|\mathcal{C}|}\mathrm{Err}(\eta_{\w,b}; \mathcal{C})
+   \lambda \tau, ~~~ \tau=\sum_{d\in D}  |w(d)| +  |b |, \]
 where $\lambda \in \R$ is a given nonnegative number.

Given a data set $\mathcal{C}$, a set $D$ of descriptors and a real $\lambda>0$, 
let LLR$(\mathcal{C},D,\lambda)$ be a procedure that returns 
a hyperplane $(\w\in \R^D,b)$  that minimizes the above function.

We review a recent learning method, called
{\em adjustive linear regression}, that is effectively
equivalent to an ANN  with no hidden layers by a linear regression
such that each input node may have a non-linear activation function
(see  \cite{ALR_ZHNA22} for the details of the idea). 
 Let $\mathcal{C}=\{\C_1,\C_2,\ldots,\C_m\}$,  $A=\{a_i=f(\C_i)\mid i\in[1,m]\}$ 
 and $X=\{x_i=f(\C_i)\in \R^D\mid i\in[1,m]\}$. 
Let $D^+$ (resp., $D^-$) denote the set of descriptor $d\in D$
such that  the correlation coefficient $\sigma(X[d],A)$ 
between $X[d]=\{x_i(d)\mid i\in[1,m]\}$ and  $A$ 
is nonnegative (resp., negative).
We first solve the following minimization problem
with a constant $\lambda\geq 0$, 
a real variable $b$
and nonnegative real variables
 $c_q(0), q\in [0,2]$, $w_q(d), q\in[0,2], d\in D$.
 
 \smallskip
\noindent 
{\bf Adjustive Linear Regression}$(\mathcal{C},\lambda)$
\begin{equation} \label{eq:minimization}
 \begin{array}{l }
\mbox{Minimize: ~}   
 \displaystyle{  \frac{1}{2m}\sum_{i\in[1,m]}
 \Bigl|c_0(0) a_i+c_1(0)  a_i^2+c_2(0)(1-(a_i\!-\! 1)^2)  } \\
    \displaystyle{ ~~~~~~~~~~~~~~~~~~~~~~~~~~~~~
         -  \sum_{d\in D^+}[ \w_0(d) \x_i(d)+ \w_1(d) \x_i(d)^2 +\w_2(d)(1-(\x_i(d)\!-\! 1)^2)]  } \\
    \displaystyle{ ~~~~~~~~~~~~~~~~~~~~~~~~~~~~~
      +  \sum_{d\in D^-}[ \w_0(d) \x_i(d)+ \w_1(d) \x_i(d)^2 + \w_2(d)(1- (\x_i(d)\!-\! 1)^2)]  - b \Bigr| +\lambda\tau } \\
\mbox{subject to}  \\ ~~~~~~~~~~~~~~~~~~~~~ 
 \displaystyle{   \tau = \sum_{d\in D} \w_0(d)   +  |b|,  ~~~
  c_0(0)+c_1(0)+c_2(0)=1.}
\end{array} 
\end{equation} 
An optimal solution to this minimization can be found by solving
a linear program with $O(m+|D|)$ variables and constraints.
From an optimal solution, 
we next compute the following hyperplane $(\w^*,b^*)$ 
to obtain a linear prediction function $\eta_{\w^*,b^*}$. 
Let
 $c^*_q(0), q\in [0,2]$, $w^*_q(d), q\in[0,2], d\in D$ and $b^*$ denote
 the values of variables 
 $c_q(0),  q\in [0,2]$, $w_q(d), q\in[0,2],  d\in D$ and $b$ 
 in an optimal solution, respectively.
 Let  $D^\dagger$ denote the set of descriptors $d\in D$ with $\w^*_0(d)>0$.
 Then we set \\
~~~ $\w^*(d):=0$ for $d\in D$ with $\w^*_0(d)=0$, \\
~~~ $\w^*(d):=\w^*_0(d)/(\w^*_0(d)+\w^*_1(d)+\w^*_2(d))$ for $ d\in D^+\cap D^\dagger$, \\
~~~ $\w^*(d):= - \w^*_0(d)/(\w^*_0(d)+\w^*_1(d)+\w^*_2(d))$ for $ d\in D^-\cap D^\dagger$ and \\
~~~  $\w^*:=(\w^*_0(1),  \w^*_0(2), \ldots, \w^*_0(|D|))\in \R^D$. 

 

\begin{figure}[h!] \begin{center}
\includegraphics[width=.80\columnwidth]{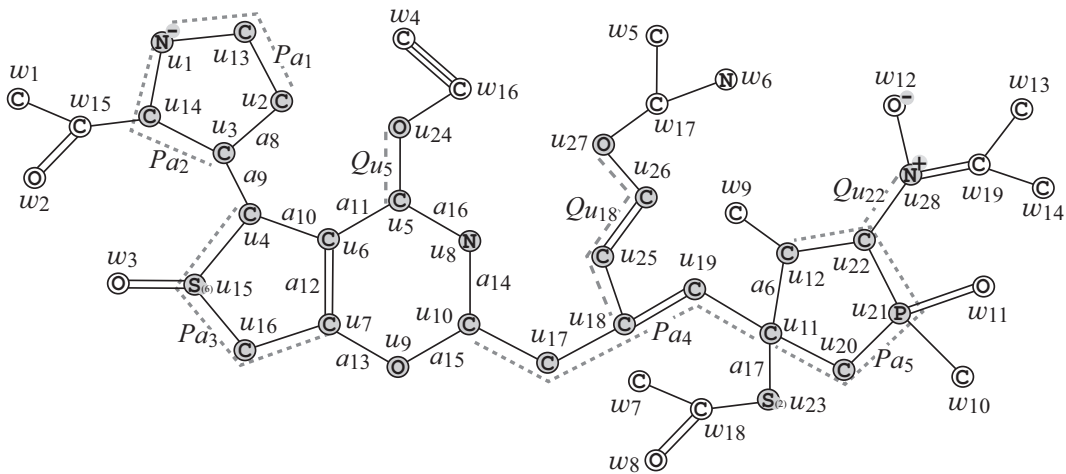}
\end{center} \caption{An illustration of  a hydrogen-suppressed chemical graph  
$\anC$ obtained from a chemical graph $\C$  by removing all the 
 hydrogens, 
where for  ${\rho}=2$,  
$V^\ex(\C)=\{w_i \mid i\in [1,19]\}$ and
$V^\inte(\C)=\{u_i \mid i\in [1,28]\}$.  
 }
\label{fig:example_chemical_graph} \end{figure}

\section{Two-layered Model}\label{sec:2LM}
This section reviews the two-layered model introduced  by 
 Shi~et~al.~\cite{SZAHZNA21}. 

 Let  $\C=(H,\alpha,\beta)$ be a chemical graph
 and  ${\rho}\geq 1$ be an integer, which we call a {\em branch-parameter}.
 
  A {\em two-layered model} of $\C$ is a partition of
 the hydrogen-suppressed chemical graph $\anC$ into
 an ``interior'' and an ``exterior'' in the following way. 
 We call a vertex $v\in V(\anC)$
   (resp., an edge $e\in E(\anC))$ of   $\C$
   an {\em exterior-vertex} (resp.,    {\em exterior-edge}) if
    $\h(v)< {\rho}$ (resp., $e$ is incident to an  exterior-vertex)
and denote the sets of exterior-vertices and exterior-edges 
by $V^\ex(\C)$ and $E^\ex(\C)$, respectively
and denote  $V^\inte(\C)=V(\anC)\setminus  V^\ex(\C)$ and 
$E^\inte(\C)=E(\anC)\setminus E^\ex(\C)$, respectively.
We call a vertex in $V^\inte(\C)$ (resp.,   an edge in $E^\inte(\C)$) 
   an {\em interior-vertex} (resp.,    {\em interior-edge}). 
 The set  $E^\ex(\C)$ of  exterior-edges forms 
a collection of connected graphs each of which is
regarded as a rooted tree $T$ rooted at 
the vertex $v\in V(T)$ with the maximum $\h(v)$. 
Let $\mathcal{T}^\ex(\anC)$ denote 
the set of these chemical rooted trees in $\anC$. 
The {\em interior} $\C^\inte$ of $\C$ is defined to be the subgraph
 $(V^\inte(\C),E^\inte(\C))$ of $\anC$. 

Figure~\ref{fig:example_chemical_graph}
 illustrates an example of a hydrogen-suppressed chemical graph $\anC$.
For a branch-parameter ${\rho}=2$, 
the interior of  the chemical graph $\anC$ in Figure~\ref{fig:example_chemical_graph} 
is obtained by removing the set of vertices with degree 1 ${\rho}=2$ times; i.e., 
first remove  
the set  $V_1=\{w_1,w_2,\ldots,w_{14}\}$ of vertices of degree 1 in $\anC$ 
and then remove  the set
 $V_2=\{w_{15},w_{16},\ldots,w_{19}\}$ of vertices of degree 1 in $\anC-V_1$,
 where the removed vertices become the exterior-vertices of $\anC$.


  
\begin{figure}[h!] \begin{center}
\includegraphics[width=.84\columnwidth]{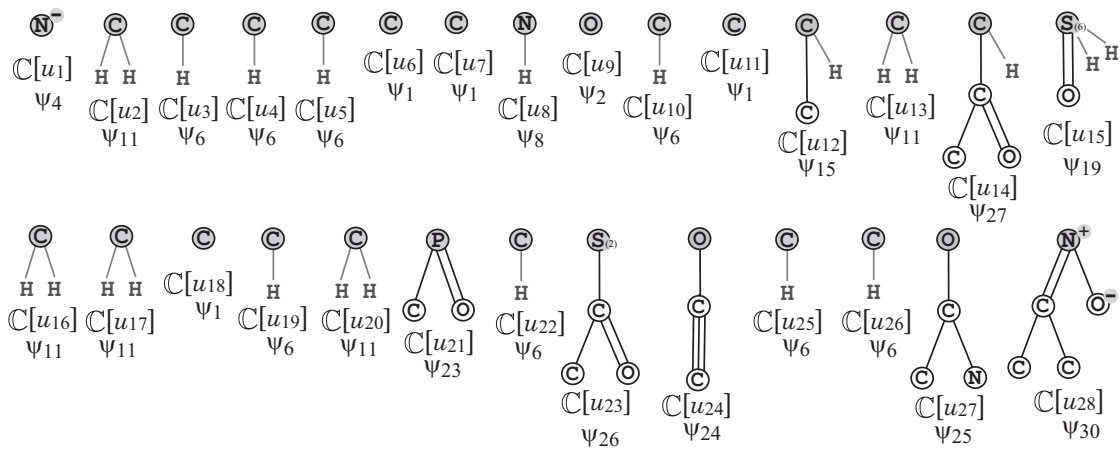}
\end{center} \caption{
The set $\mathcal{T}(\C)$ of  2-fringe-trees  $\C[u_i], i\in [1,28]$ of the example $\C$
in Figure~\ref{fig:example_chemical_graph}, 
where the root of each tree is depicted with a gray circle and
 the hydrogens attached to non-root vertices are omitted in the figure.  
 }
\label{fig:example_fringe-tree} \end{figure}

For each interior-vertex $u\in V^\inte(\C)$,
let $T_u\in \mathcal{T}^\ex(\anC)$ denote the chemical tree rooted at $u$
(where possibly $T_u$ consists of vertex $u$)
and 
define the {\em $\rho$-fringe-tree} $\C[u]$ to be  
the chemical rooted tree obtained from $T_u$ by putting back
 the hydrogens originally attached $T_u$ in $\C$. 
Let $\mathcal{T}(\C)$ denote the set of $\rho$-fringe-trees 
$\C[u], u \in V^\inte(\C)$. 
Figure~\ref{fig:example_fringe-tree}  illustrates
the set  $\mathcal{T}(\C)=\{\C[u_i]\mid i\in [1,28]\}$ of the 2-fringe-trees 
  of the example $\C$
in Figure~\ref{fig:example_chemical_graph}. 

\smallskip
\noindent {\bf Feature Function~} 
 The feature of an  interior-edge $e=uv\in E^\inte(\C)$ 
 such that $\alpha(u)=\ta$, $\deg_{\anC}(u)=d$, 
 $\alpha(v)=\tb$, $\deg_{\anC}(v)=d'$  and $\beta(e)=m$  is represented by 
 a tuple $(\ta d, \tb d', m)$, which is called the {\em edge-configuration} 
  of the edge $e$, where 
  we call the tuple $(\ta, \tb, m)$ 
 the {\em adjacency-configuration} of the edge $e$. 
 

In the framework with the two-layered model,
the feature vector $f$ mainly consists of the frequency 
of edge-configurations of   the interior-edges  and
the frequency of chemical rooted trees among the set 
of  chemical rooted trees $\C[u]$ over all interior-vertices $u$. 
See Appendix~\ref{sec:descriptor} for  all these descriptors $x(1),x(2),\ldots,x(K_1)$,
which are called {\em linear descriptors}.
We denote by  $D_\pi^{(1)}:=\{x(k)\mid k\in[1,K_1]\}$ the set of descriptors
constructed over a data set for a property $\pi$. 
In this paper, we also use a quadratic term $x(i)x(j)$ (or $x(i)(1-x(j))$), $1\leq i\leq j\leq K_1$
as a new descriptor, where we assume that each $x(i)$ is normalized between 0 and 1.
We call such a  term $x(i)x(j)$ (or $x(i)(1-x(j))$), $1\leq i\leq j\leq K_1$ 
a {\em quadratic descriptor} and denote by 
$D_\pi^{(2)}:=\{x(i)x(j)\mid 1\leq i\leq j\leq K_1\}\cup
 \{ x(i)(1-x(j))\mid 1\leq i, j\leq K_1\}$ the set of  quadratic descriptors. 

To construct a prediction function, 
we use the union   $D_\pi^{(1)} \cup D_\pi^{(2)}$. 
 This set of descriptors is usually excessive in constructing a prediction function,  and 
 we reduce it to a smaller set of descriptors to construct a {\em feature function}
 $f: \RK\to \R$, where $K$ is the  number of resulting descriptors.
 We call  $\RK$ {\em  the feature space}.
 See Appendix~\ref{sec:reduce_descriptors}  for methods of reducing descriptors.
 
 
\smallskip
\noindent {\bf Topological Specification~}   
A topological specification $\sigma$ is described
as a set of the following rules:
\begin{enumerate}[nosep]
\item[(i)]
a {\em seed graph} $\GC$ as an  abstract form of  a target chemical graph $\C$;
\item[(ii)]
 a set $\mathcal{F}$ of chemical rooted trees  as candidates
 for a tree  $\C[u]$ rooted at each exterior-vertex $u$ in $\C$; 
and 
\item[(iii)]
lower and upper bounds on the number of components 
 in a target chemical graph such as  chemical elements, 
double/triple bonds and the interior-vertices in $\C$. 
\end{enumerate} 

\begin{figure}[h!] \begin{center}
\includegraphics[width=.98\columnwidth]{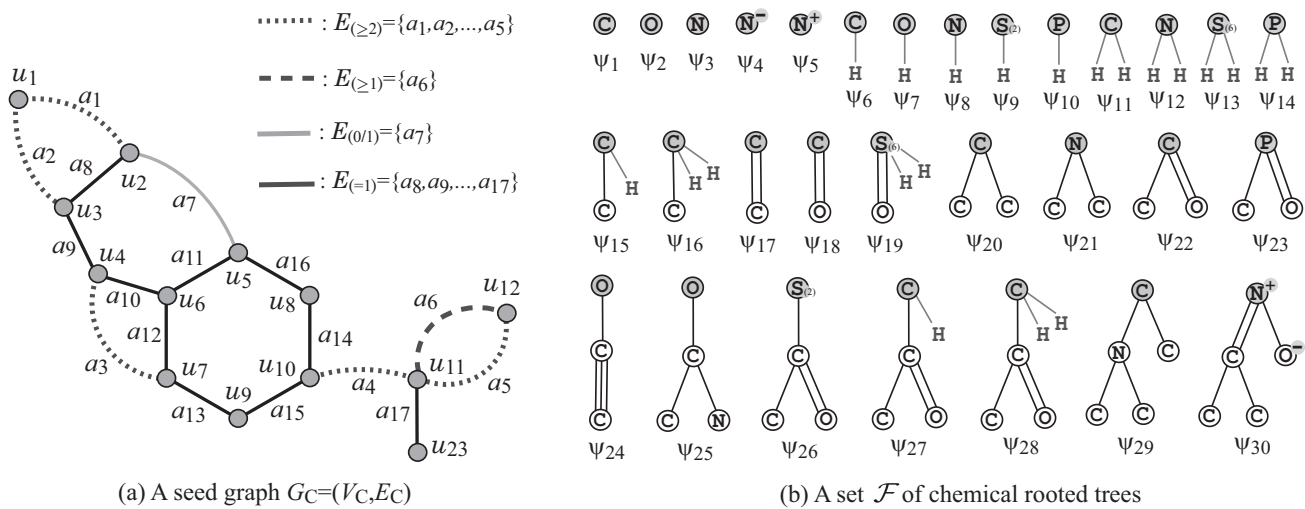}
\end{center} \caption{
(a) An illustration of a seed graph $\GC$, 
where the vertices in $\VC$ are depicted with gray circles,
the edges in $\Et$ are depicted with dotted lines,
the edges in $\Ew$ are depicted with dashed lines,
the edges in $\Ez$ are depicted with gray bold lines and  
the edges in $\Eew$ are depicted with black solid lines;
(b) A set $\mathcal{F}=\{\psi_1,\psi_2,\ldots,\psi_{30}\}\subseteq
\mathcal{F}(\mathcal{C}_\pi)$ of 30 chemical rooted trees
$\psi_i, i\in [1,30]$, where the root of each tree is depicted with a gray circle.
The hydrogens attached to non-root vertices are omitted in the figure.    }
\label{fig:specification_example_1} \end{figure}  

Figures~\ref{fig:specification_example_1}(a) and (b)
 illustrate  examples of  a  seed graph  $\GC$ and 
 a set $\mathcal{F}$ of chemical rooted trees, respectively. 
 Given a seed graph $\GC$, 
 the interior of   a target chemical graph $\C$ is constructed
 from $\GC$ by replacing some edges $a=uv$ 
 with paths $P_a$ between the end-vertices
 $u$ and $v$ and by attaching new paths $Q_v$ to some vertices $v$.  
%
For example, a chemical graph $\C$ 
in Figure~\ref{fig:example_chemical_graph} is constructed
from the seed  graph  $\GC$ in Figure~\ref{fig:specification_example_1}(a)
as follows.
\begin{enumerate}[nosep,  leftmargin=*]
\item[-]
First replace  five edges
 $a_1=u_1 u_{2},  a_2=u_1 u_{3},  a_3=u_4 u_{7}, a_4=u_{10}u_{11}$
and $a_5=u_{11}u_{12}$ in  $\GC$ 
 with new paths  
$P_{a_1}=(u_1,u_{13},u_{2})$, 
$P_{a_2}=(u_{1},u_{14},u_{3})$,
$P_{a_3}=(u_{4},u_{15},u_{16},u_{7})$, 
 $P_{a_4}=(u_{10},u_{17},u_{18},u_{19},u_{11})$ and
 $P_{a_5}=(u_{11},u_{20},u_{21},u_{22},u_{12})$, respectively
 to obtain a subgraph $G_1$ of $\anC$. 
\item[-]
Next attach to this graph  $G_1$ three new paths 
$Q_{u_5}=(u_5,u_{24})$, 
$Q_{u_{18}}=(u_{18},u_{25},u_{26},u_{27})$ and 
$Q_{u_{22}}=(u_{22},u_{28})$
to obtain  
the interior of  $\anC$ in Figure~\ref{fig:example_chemical_graph}.
\item[-]
Finally  attach to the interior   28 trees selected from the set $\mathcal{F}$ 
and assign chemical elements and bond-multiplicities in the interior
to  obtain a chemical graph $\C$  in Figure~\ref{fig:example_chemical_graph}. 
In Figure~\ref{fig:example_fringe-tree},  
  $\psi_1\in \mathcal{F}$ is selected for $\Co[u_i]$, $i\in\{6,7,11\}$.
    Similarly 
  $\psi_2$  for  $\Co[u_9]$,
  $\psi_4$   for $\Co[u_1]$, 
  $\psi_6$   for $\Co[u_i]$,   
  $i\in\{3,4,5,10,19,22,25,26\}$,
  $\psi_8$   for $\Co[u_8]$, 
  $\psi_{11}$    for $\Co[u_i]$, $i\in\{2,13,16,17,20\}$,
  $\psi_{15}$   for $\Co[u_{12}]$,
   $\psi_{19}$    for $\Co[u_{15}]$,
   $\psi_{23}$    for $\Co[u_{21}]$,
   $\psi_{24}$    for $\Co[u_{24}]$,
   $\psi_{25}$    for $\Co[u_{27}]$, 
   $\psi_{26}$   for $\Co[u_{23}]$, 
   $\psi_{27}$  for $\Co[u_{14}]$     
   and 
    $\psi_{30}$  for $\Co[u_{28}]$. 
\end{enumerate} 

%
%

See Appendix~\ref{sec:specification} for a full description of topological specification.

\section{How to Compute a Quadratic Term in an MILP}\label{sec:compute_Q}

This section introduces an MILP formulation for computing the product
of two descriptors in an MILP.

Given two real values $x$ and $y$ with $0\leq x\leq 1$ and $0\leq y\leq 1$, 
the computing process of the product $z=xy$
can be approximately formulated as the following MILP.
First regard $(2^{p+1}-1) x$ as an integer with a binary expression of $p+1$ bits,
where $x^{(j)}\in [0,1]$ denotes the value of the $j$-th bit.
Then compute  $y\cdot  x^{(j)}$ which becomes 
the  $j$-th bit $z^{(j)}$ of  $(2^{p+1}-1)z$.

\smallskip\noindent
{\bf constants: } 
\begin{enumerate}[leftmargin=*]
\item[-]  
 $x,y$: reals with $0\leq x, y \leq 1$; 
 
\item[-]  
$p$: a positive integer;

\end{enumerate}

\smallskip\noindent
{\bf variables: }   
\begin{enumerate}[leftmargin=*]

\item[-]
 $z$, $z^{(j)}, j\in[0,p]$: reals with $0\leq z, z^{(j)}\leq 1$; 
 
\item[-]
 $x^{(j)}\in [0,1], j\in[0,p]$: binary variables; 

\end{enumerate}

\smallskip\noindent
{\bf constraints: }    
\begin{align}    
 \sum_{j\in[0,p]}2^j x^{(j)}-1 \leq  
(2^{p+1}-1) x\leq \sum_{j\in[0,p]}2^j x^{(j)},  && \notag \\
z^{(j)}\leq x^{(j)}, && j\in[0,p], \notag \\
y - (1-x^{(j)}) \leq z^{(j)} \leq y + (1-x^{(j)}), && j\in[0,p], \notag \\
z =\frac{1}{2^{p+1}-1}\sum_{ j\in[0,p]} 2^j z^{(j)}.  \label{ej:ex_1}  
\end{align}    

Note that the necessary number of integer variables for computing
$xy$ for one pair of $x$ and $y$ is $p$.
In this paper, we set $p:=6$ in our computational experiment.
The relative error by $p=6$  in the above method is  at most 
 $\frac{1}{2^{p+1}-1}=1/127$,
 which is around $0.8\%$.


\section{Results}\label{sec:experiment}

With our new method of choosing descriptors and
formulating an MILP to treat quadratic descriptors in the two-layered model,
we implemented the framework
for inferring chemical graphs  and
conducted experiments  to evaluate the computational efficiency. 
We executed the experiments on a PC with 
 Processor:  Core i7-9700 (3.0GHz; 4.7 GHz at the maximum) and 
Memory: 16 GB RAM DDR4. 
To construct a prediction function by LLR, MLR or ANN, 
we used {\tt scikit-learn} version 1.0.2  with Python 3.8.12, 
MLPRegressor and ReLU activation function. 

\subsection{Results on the First Phase of the Framework}

\noindent
 {\bf Chemical properties }
We implemented the first phase for the following  
32 chemical properties of monomers and
ten chemical properties of  polymers.
 
For monomers, 
 we used the following data sets: \\ 
~  biological half life ({\sc BHL}), 
    boiling point  ({\sc Bp}),
     critical temperature ({\sc Ct}),
     critical pressure  ({\sc Cp}),  \\
~   dissociation constants ({\sc Dc}),   
     flash point in closed cup ({\sc Fp}),  
     heat of combustion ({\sc Hc}),  \\
~   heat of vaporization  ({\sc Hv}), 
     octanol/water partition coefficient  ({\sc Kow}),  
     melting point  ({\sc Mp}),  \\
~   optical rotation  ({\sc OptR}),
     refractive index of trees ({\sc RfIdT}),
      vapor density   ({\sc Vd}) and  \\
~   vapor pressure    ({\sc Vp}), 
 provided  by HSDB from PubChem~\cite{pubchem}; \\  
~ electron density on the most positive atom ({\sc EDPA})
and 
Kovats retention index ({\sc Kov})  \\
~  by  M.~Jalali-Heravi and M.~Fatemi~\cite{JF01};  \\ 
~ entropy ({\sc ET}) by P.~Duchowicz et al.~\cite{DCT02}; \\ 
~ heat of atomization ({\sc Ha}) and heat of formation ({\sc Hf})
by K.~Roy and A.~Saha~\cite{RS03}; \\ 
~ surface tension ({\sc SfT}) by V.~Goussard et al.~\cite{GFPDNA17};  \\ 
~  viscosity  ({\sc Vis}) by   V.~Goussard et al.~\cite{GFPDNA20}; \\ 
~ isobaric heat capacities liquid ({\sc LhcL}) and 
isobaric heat capacities solid ({\sc LhcS})
by R.~Naef~\cite{Naef}; \\
~  lipophilicity  ({\sc Lp}) by N.~Xiao~\cite{figshare}; \\ 
~ flammable limits lower of organics ({\sc FlmLO}) 
 by S.~Yuan et al.~\cite{YJQKM19}; \\ 
~ 
molar refraction at 20 degree ({\sc Mr})
by Y.~M.~Ponce~\cite{Ponce03}; and \\
~ solubility ({\sc Sl}) by ESOL~\cite{moleculenet},\\ 
~ energy of highest occupied molecular orbital ({\sc Homo}),  \\
~  energy of lowest unoccupied molecular orbital ({\sc Lumo}),   \\
~  the energy difference between  {\sc Homo} and {\sc Lumo} ({\sc Gap}),    \\
~  isotropic polarizability ({\sc Alpha}),
heat capacity at 298.15K ({\sc Cv}),   
internal energy at 0K ({\sc U0})  \\
~  and 
electric dipole moment ({\sc mu})   provided by ESOL~\cite{moleculenet},
where the properties  from {\sc Homo} \\
~ to {\sc mu}   are based on a common data set QM9.

 The data set  QM9 contains more than 130,000 compounds.
 In our experiment, we use a set of  1,000 compounds randomly selected from the data set. 
For property {\sc Hv},  we remove the chemical compound with CID=7947
as an outlier  from the original data set.   

For polymers, we used the following data  provided by  J.~Bicerano~\cite{Bicerano}: \\
~  experimental amorphous density ({\sc AmD}), 
   characteristic ratio ({\sc ChaR}),\\
~ dielectric constant ({\sc DieC}),  dissipation factor ({\sc DisF}),
   heat capacity in liquid ({\sc HcL}),    \\
~  heat capacity in solid ({\sc HcS}), mol volume ({\sc MlV}),  permittivity ({\sc Prm}), \\
~  refractive index of polymers ({\sc RfIdP})   and  glass transition ({\sc Tg}), \\
\noindent
 where 
 we  excluded from our test data set every polymer  whose chemical 
 formula  could not be found by its name in the book~\cite{Bicerano}.  
We remark that the previous learning experiments for $\pi\in \{${\sc ChaR}, {\sc RfIdP}$\}$
based on the two-layered model due to
Azam~et~al.~\cite{gridAZHZNA21} and Zhu~et~al.~\cite{ALR_ZHNA22}
excluded some number of polymers as outliers.
In our experiments, we do not exclude any polymer from the original data set as outliers
 for these properties.

\begin{table}[h!]\caption{Results of setting data sets for monomers.} 
  \begin{center}
    \begin{tabular}{@{} c c r c  c  r r r   @{}}\hline
      $\pi$ & $\Lambda$  &  $|\mathcal{C}_{\pi}|$  & 
       $ \underline{n},~\overline{n} $ &   $\underline{a},~\overline{a}$ &
   $|\Gamma|$   &  $|\mathcal{F}|$ &   $K_1$ \\ \hline 
      {\sc BHL} & $\Lambda_7$  & 514  &  5,\,36  &  0.03,\,732.99   & 26  & 101  & 166    \\
      {\sc Bp} & $\Lambda_2$  & 370  &  4,\,67  &  -11.7,\,470.0   & 22   & 130  & 184      \\
      {\sc Bp} & $\Lambda_7$  & 444  &  4,\,67  &  -11.7,\,470.0   & 26   & 163 & 230     \\ 
      {\sc Cp} & $\Lambda_5$  & 131  &  4,\,63  &  $4.7\!\times\! 10^{-6}$,\,5.52  & 8   & 79  & 119       \\ 
       {\sc Ct} & $\Lambda_2$  &  125 &  4,\,63 &   56.1,\,3607.5 &  8  &  76  & 113 \\
       {\sc Ct} & $\Lambda_5$  &  132 &  4,\,63 &   56.1,\,3607.5  &  8  & 81   & 121 \\
      {\sc Dc} & $\Lambda_2$  & 141  &  5,\,44  &  0.5,\,17.11   & 20   & 62  & 111     \\ 
      {\sc Dc} & $\Lambda_7$  & 161  &  5,\,44  &  0.5,\,17.11   & 25   & 69  & 130     \\  
      {\sc ET} & $\Lambda_7$    & 17  &  5,\,12  & 64.34,\,96.21   &  5  &  17   &  53   \\  
      {\sc Fp} & $\Lambda_2$  & 368  &  4,\,67  &  -82.99,\,300.0   & 20   & 131  & 183      \\ 
      {\sc Fp} & $\Lambda_7$  & 424  &  4,\,67  &  -82.99,\,300.0   & 25   & 161  & 229       \\   
      {\sc FlmLO} & $\Lambda_{16}$   & 1046   & 1,\,49   & 0.185,\,4.3  &  34   & 282   & 376 \\  
       {\sc Hv} & $\Lambda_2$  &   94  &  4,\,16 &  19.12,\,210.3 & 12   & 63   & 105 \\
      {\sc Kov} & $\Lambda_1$  & 52  &  11,\,16  &  1422.0,\,1919.0   & 9   & 33  & 64      \\ 
      {\sc Kow} & $\Lambda_2$  & 684  &  4,\,58  &  -7.5,\,15.6  & 25   & 166  & 223 \\
    {\sc Kow} & $\Lambda_8$  & 899 & 4,\,69 & -7.5,\,15.6 & 37   & 219    & 303 \\       
      {\sc Lp} & $\Lambda_2$  & 615  &  6,\,60  &  -3.62,\,6.84   & 32   & 116  & 186      \\ 
      {\sc Lp} & $\Lambda_8$  & 936  &  6,\,74  &  -3.62,\,6.84  & 44   & 136  & 231     \\ 
      {\sc Mp} & $\Lambda_2$ & 467 &4,\,122 &   -185.33,\,300.0  & 23 &142   & 197 \\
      {\sc Mp} & $\Lambda_8$ & 577 &4,\,122 &   -185.33,\,300.0  & 32 &176   & 255 \\ 
      {\sc OptR} & $\Lambda_2$  & 147  & 5,\,44  &  -117.0,\,165.0   & 21   & 55  & 107      \\ 
      {\sc OptR} & $\Lambda_4$  & 157  &  5,\,69  &  -117.0,\,165.0   & 25   & 62  & 123     \\ 
      {\sc RfIdT} &   $\Lambda_{10}$  & 191  & 4,\,26 &  0.919,\,1.613 &  17 &  115 &  168 \\
      {\sc Sl} & $\Lambda_2$  & 673  &  4,\,55  &  -9.332,\,1.11   & 27   & 154  & 217     \\ 
      {\sc Sl} & $\Lambda_8$  & 915 &  4,\,55  &  -11.6,\,1.11  & 42   & 207  & 300      \\ 
      {\sc SfT} & $\Lambda_3$  & 247  &  5,\,33  & 12.3,\,45.1  & 11   & 91  & 128      \\     
      {\sc Vis} & $\Lambda_3$  & 282  &  5,\,36  &  -0.64,\,1.63   & 12   & 88  & 126      \\ 
      {\sc Homo} &  $\Lambda_9$  & 977 &  6,\,9   &-0.3335,\, -0.1583  & 59  & 190  & 297 \\
      {\sc Lumo} & $\Lambda_9$  & 977  &  6,\,9  &  -0.1144,\,0.1026   & 59   & 190 & 297   \\
      {\sc Gap} & $\Lambda_9$  &  977  &  6,\,9   & 0.1324,\,0.4117  & 59 &  190  & 297 \\
      {\sc Alpha} & $\Lambda_9$ & 977   &  6,\,9     & 50.9,\,99.6 & 59 & 190  & 297  \\ 
       {\sc Cv} &  $\Lambda_9$ &  977  &6,\,9   &    19.2,\,44.0 & 59  & 190    & 297  \\
       {\sc mu} & $\Lambda_9$ & 977 & 6,\,9   &   0.04,\,6.897   & 59  &190   & 297 \\   
      \hline
  \end{tabular}\end{center}\label{table:phase1a2}
\end{table} 

\begin{table}[h!]\caption{Results of setting data sets for polymers.} 
  \begin{center}
    \begin{tabular}{@{} c c r c  c  r r r   @{}}\hline
      $\pi$ & $\Lambda$  &  $|\mathcal{C}_{\pi}|$  &  $ \underline{n},~\overline{n} $ &  
       $\underline{a},~\overline{a}$ &   $|\Gamma|$   &  $|\mathcal{F}|$ &   $K_1$ \\ \hline
      {\sc AmD} & $\Lambda_2$  & 86  &  4,\,45  & 0.838,\,1.34 & 16  & 25    & 83  \\ 
      {\sc AmD} &  $\Lambda_{13}$ &  93 &   4,\,45  & 0.838,\,1.45 & 18   &  30  & 94  \\
       {\sc ChaR} &  $\Lambda_2$  & 30   &  4,\,18   &  3.7,\,15.9  &   15   &  17   &  68 \\ 
      {\sc ChaR} & $\Lambda_{12}$  & 32  & 4,\,18  & 3.7,\,15.9  & 15  & 18 &  71 \\ 
      {\sc ChaR} & $\Lambda_6$  &  35  & 4,\,18  & 3.7,\,15.9  & 18  & 21 &  83 \\     
      {\sc DeiC} &  $\Lambda_{12}$  & 36 &  4,\,22  & 2.13,\,3.4  & 11  &18   & 67   \\
      {\sc DisF} &  $\Lambda_{13}$ &  132 &  4,\,45& $7\!\times\! 10^{-5}$,\,0.07 &  15 &  18  & 78   \\  
      {\sc Prm} & $\Lambda_2$ &  112  &  4,\,45  & 2.23,\,4.91  & 14 &   15  & 69    \\ 
      {\sc Prm} &  $\Lambda_{13}$  & 132 &  4,\,45  & 2.23,\,4.91  & 15  &18   & 78  \\
      {\sc RfIdP} & $\Lambda_{11}$ & 92 & 4,\,29 &  0.4899,\,1.683  & 15  & 35  & 96    \\  
      {\sc RfIdP} & $\Lambda_{14}$  & 125 & 4,\,29 &  0.4899,\,1.683  & 19  & 50  & 124  \\  
      {\sc RfIdP} & $\Lambda_{15}$  & 135 & 4,\,29 &  0.4899,\,1.71 & 23  & 56  & 144      \\  
       {\sc Tg} & $\Lambda_2$  & 204  &  4,\,58  &    171,\,673 & 19   & 36 & 101 \\
      {\sc Tg} & $\Lambda_7$  & 232  & 4,\,58 &   171,\,673 & 21   & 43  & 118 \\
      \hline
  \end{tabular}\end{center}\label{table:phase1a3}
\end{table}

 \medskip \noindent
 {\bf Setting data sets }
For each property $\pi$,  
 we first select a set $\Lambda$ of chemical elements 
 and then collect  a  data set  $\mathcal{C}_{\pi}$ on chemical graphs
 over the set $\Lambda$ of chemical elements.  
 To construct the data set $\mathcal{C}_{\pi}$,
  we eliminated  chemical compounds that do not satisfy 
  one of the following: the graph is connected,
  the number of carbon atoms is at least four,
  and   the number of non-hydrogen neighbors of each atom is
  at most 4.   

We set a branch-parameter ${\rho}$ to be 2, 
introduce linear   descriptors defined by the two-layered graph 
in the chemical model without suppressing hydrogen
and use the set $D_\pi^{(1)}\cup D_\pi^{(2)}$ 
of linear and quadratic descriptors  
(see Appendix~\ref{sec:descriptor} for the details).
 We normalize the range of each linear descriptor and
 the range $\{t\in \R \mid \underline{a}\leq t\leq \overline{a}\}$ 
 of property values   $a(\Co), \Co\in \mathcal{C}_\pi$.

 We compare the following four methods of constructing a prediction function.
 \begin{enumerate}
 \item[(i)] {\bf LLR}: use Lasso linear regression on the  set  the  $D_\pi^{(1)}$ of linear descriptors
 (see \cite{ZAHZNA21} for the detail of the implementation); 
 \item[(ii)]  {\bf ANN}:  use ANN on the  set  the  $D_\pi^{(1)}$ of linear descriptors
 (see \cite{ZAHZNA21} for the detail of the implementation); 
 \item[(iii)]  {\bf ALR}:  use adjustive linear regression on the  set  the  $D_\pi^{(1)}$ of linear descriptors
 (see \cite{ALR_ZHNA22} for the detail of the implementation); and
 \item[(iv)]  {\bf R-MLR}:  apply our method  (see Appendix~\ref{sec:reduce_descriptors})
  of reducing descriptors to 
 the set $D_\pi^{(1)}\cup D_\pi^{(2)}$ of linear and quadratic descriptors and
  use multi-linear regression for the resulting set of descriptors. 
\end{enumerate}

 Among the above properties,   we found that 
 the median of test  coefficient of determination ${\rm R}^2$ 
 of the prediction function constructed 
  by LLR~\cite{ZAHZNA21} or ALR~\cite{ALR_ZHNA22} 
 exceeds 0.98 for the following 
   nine properties of monomers (resp., three properties of polymers):  
     {\sc EDPA},  {\sc Hc},  {\sc Ha}, {\sc Hf}, {\sc LhcL},  {\sc LhcS}, {\sc Mr}, {\sc Vd} and  {\sc U0}   
 (resp.,  {\sc HcL}, {\sc HcS} and {\sc MlV}). 
We excluded the above properties in the following experiment, 
and used the rest of 
23 
chemical properties of monomers and
seven 
chemical properties of polymers to compare the four methods (i)-(iv).

 Tables~\ref{table:phase1a2}  and  \ref{table:phase1a3}   show 
  the size and range of data sets   that 
 we prepared for each chemical property to construct a prediction function,
 where  we denote the following:  
\begin{enumerate}[nosep,  leftmargin=*]
\item[-] $\pi$: the name of a chemical property used in the experiment.
\item[-] 
  $\Lambda$: a set of selected elements used in the data set $\mathcal{C}_{\pi}$; 
  $\Lambda$ is one of the following 19 sets:  \\
  $\Lambda_1=\{\ttH,\ttC,\ttO \}$; 
   $\Lambda_2=\{\ttH,\ttC,\ttO, \ttN \}$;
   $\Lambda_3=\{\ttH,\ttC,\ttO, \ttSi_{(4)} \}$;  
     $\Lambda_4=\{\ttH,\ttC,\ttO, \ttN,\ttS_{(2)},\ttF \}$; \\
      $\Lambda_5=\{\ttH,\ttC,\ttO, \ttN,  \ttCl, \ttPb \}$;
    $\Lambda_6=\{\ttH,\ttC,\ttO, \ttN,\ttSi_{(4)},\ttCl,\ttBr \}$;   
   $\Lambda_7=\{\ttH,\ttC,\ttO, \ttN,\ttS_{(2)},\ttS_{(6)},\ttCl \}$;     \\  
   $\Lambda_8=\{\ttH,\ttC,\ttO, \ttN,\ttS_{(2)},\ttS_{(4)},\ttS_{(6)},\ttCl \}$;  
    $\Lambda_9=\{\ttH, \ttC_{(2)},\ttC_{(3)},\ttC_{(4)},\ttC_{(5)},\ttO,
   \ttN_{(1)}, \ttN_{(2)}, \ttN_{(3)}, \ttF \}$; \\
  $\Lambda_{10}=\{\ttH,\ttC,\ttO,  \ttN,\ttP_{(2)},\ttP_{(5)},\ttCl \}$; 
 $\Lambda_{11}=\{\ttH, \ttC, \ttO_{(1)}, \ttO_{(2)}, \ttN \}$;  
   $\Lambda_{12}=\{\ttH, \ttC, \ttO, \ttN, \ttCl \}$;  \\
   $\Lambda_{13}=\{\ttH, \ttC, \ttO, \ttN,  \ttCl, \ttS_{(2)} \}$;  
   $\Lambda_{14}=\{\ttH, \ttC, \ttO_{(1)}, \ttO_{(2)}, \ttN,  \ttCl, \ttSi_{(4)}, \ttF \}$;      \\ 
  $\Lambda_{15}=\{\ttH,\ttC,\ttO_{(1)}, \ttO_{(2)}, \ttN, \ttSi_{(4)},\ttCl,\ttF,
                            \ttS_{(2)}, \ttS_{(6)},\ttBr \}$;   \\ 
%
   $\Lambda_{16}=\{\ttH, \ttC, \ttO_{(2)}, \ttN,  \ttCl, \ttP_{(3)}, \ttP_{(5)},
    \ttS_{(2)}, \ttS_{(4)}, \ttS_{(6)}, \ttSi_{(4)}, \ttBr , \ttI \}$, 
 where ${\tt a}_{(i)}$ for a chemical element ${\tt a}$ and an integer $i\geq 1$ 
 means that  a chemical element ${\tt a}$ with valence $i$. 

\item[-] 
 $|\mathcal{C}_{\pi}|$:  the size of data set $\mathcal{C}_{\pi}$ over $\Lambda$
  for the property $\pi$.
   
\item[-]   $ \underline{n},~\overline{n} $:  
  the minimum and maximum  values of the number 
  $n(\Co)$ of non-hydrogen atoms in 
  the   compounds $\Co$ in $\mathcal{C}_{\pi}$.
\item[-] $ \underline{a},~\overline{a} $:  the minimum and maximum values
of $a(\Co)$ for $\pi$ over   the   compounds $\Co$ in  $\mathcal{C}_{\pi}$.
\item[-]    $|\Gamma|$: 
the number of different edge-configurations
of interior-edges over the compounds in~$\mathcal{C}_{\pi}$. 
\item[-]  $|\mathcal{F}|$: the number of non-isomorphic chemical rooted trees
 in the set of all 2-fringe-trees in  the   compounds in $\mathcal{C}_{\pi}$.
 
\item[-]  $K_1$: the size  $|D_\pi^{(1)}|$ of a set $D_\pi^{(1)}$ of linear descriptors,
where $|D_\pi^{(2)}|=(3(K_1)^2+K_1)/2$ holds.  
\end{enumerate}

\medskip \noindent
{\bf Constructing prediction functions }
For each chemical property $\pi$, we construct a prediction function
by one of the four methods (i)-(iv). 

For methods (i)-(iii), we used the same implementation  
due to Zhu~et al.~\cite{ZAHZNA21,ALR_ZHNA22}
and omit the details.

In method  (iv), we use our new 
 procedures LLR-Reduce and Select-Des-set for reducing the number of descriptors
  (see Appendix~\ref{sec:reduce_descriptors} for the details). 
Method (iv) for property $\pi$ is implemented as follows. 
If $\pi$ is a monomer property and $|D_\pi^{(1)}\cup D_\pi^{(2)}|>5000$ then first  execute
LLR-Reduce$(\mathcal{C}_\pi, D_\pi^{(1)}\cup D_\pi^{(2)})$
to find a subset $\widetilde{D}$ of $D_\pi^{(1)}\cup D_\pi^{(2)}$ with $5000$ descriptors.
Otherwise set  $\widetilde{D}:=D_\pi^{(1)}\cup D_\pi^{(2)}$. 
Next execute Select-Des-set$(\mathcal{C}_\pi,\widetilde{D})$ 
 to obtain a subset $D^*$ of  $\widetilde{D}$. 
Construct a prediction function by MLR  
on the selected descriptor set $D^*$.

 Tables~\ref{table:phase1b2}  and \ref{table:phase1b3} show  
 the results on constructing prediction functions,
 where  we denote the following:     
\begin{enumerate}[nosep,  leftmargin=*]  
\item[-] $\pi$: the name of a chemical property used in the experiment.

\item[-]  
  $\Lambda$: the set of elements selected from the data set $\mathcal{C}_{\pi}$. 

\item[-] LLR:  the median of test $\mathrm{R}^2$  
  in ten 5-fold cross-validations for prediction functions constructed by method (i).  

\item[-] ANN:
the median of test $\mathrm{R}^2$  
  in ten 5-fold cross-validations for prediction functions constructed by method (ii).  
  
\item[-] ALR:
the median of test $\mathrm{R}^2$  
  in ten 5-fold cross-validations for prediction functions constructed by method (iii).  

\item[-] R-MLR:
the median of test $\mathrm{R}^2$  
  in ten 5-fold cross-validations for prediction functions constructed by method (iv).  
 
\item[-] the score of LLR, ANN, ALR or R-MLR   marked with ``*'' indicates
the best performance among the four methods for the property $\pi$; 
  
\item[-] $K^*_{\mathrm{1}}, K^*_{\mathrm{2}}$: the numbers
$K^*_{\mathrm{1}}$ and $K^*_{\mathrm{2}}$ of linear and quadratic descriptors, respectively
 in the set $D^*$ selected by our method (iv)
 from the set $D_\pi^{(1)}\cup D_\pi^{(2)}$ 
 before a prediction function is constructed by MLR in (iv).
\end{enumerate}
The running time of choosing a descriptor set $D^*$ in method  (iv) was  around
from  80 to  $4\times 10^4$ seconds and the time for constructing a prediction function
to $D^*$ is around 0.03 to 0.46 second. 
  

\begin{table}[h!]\caption{Results of constructing prediction functions for monomers.} 
  \begin{center}
    \begin{tabular}{@{} c l    r  r r r   r   c c c c  @{}}\toprule
      $\pi$ & ~$\Lambda$  &       LLR & ANN  &    ALR    & R-MLR 
        & $K^*_{\mathrm{1}},K^*_{\mathrm{2}}$     \\ \midrule 
      {\sc BHL}    & $\Lambda_7$  &  0.483 & 0.622  &  0.265 & *0.659   &  0,\,27~\,  \\
      {\sc Bp} & $\Lambda_2$      & 0.599   & 0.765  & 0.816 & *0.935 & 1,\,59~\,   \\
      {\sc Bp} & $\Lambda_7$      &  0.663 & 0.720   & 0.832 & *0.899 &  0,\,38~\,  \\
      {\sc Cp} & $\Lambda_5$     & 0.555  &  0.727  & 0.690  &  *0.841 & 0,\,67~\, \\
      {\sc Ct} & $\Lambda_2$     &  0.037  & 0.357   & 0.900 &  *0.937 &  1,\,47~\, \\ 
      {\sc Ct} & $\Lambda_5$    & 0.048  & 0.357 & *0.895 &  0.860 &  0,\,13~\,  \\ 
      {\sc Dc}   & $\Lambda_2$   &  0.489 & 0.651  &   0.488 &  *0.908 &  0,\,58~\,  \\ 
      {\sc Dc} &  $\Lambda_7$  & 0.574  & 0.622 &  0.602 &  *0.829  & 0,\,26~\,   \\
      {\sc ET} &   $\Lambda_7$   & 0.132   &0.479   & 0.464   & *0.996   & 0,\,13~\,  \\
      {\sc Fp} & $\Lambda_2$    & 0.589 & 0.746    & 0.719 &  *0.899  & 0,\,42~\,  \\ 
      {\sc Fp} & $\Lambda_7$   &  0.571 & 0.745  &     0.684 & *0.846  &  0,\,32~\,  \\ 
      {\sc FlmLO} &   $\Lambda_{16}$   & 0.819  &  0.928 &   0.604 &   *0.949  & 0,\,77~\,   \\ 
      {\sc Hv} & $\Lambda_2$    & 0.864  & 0.778 &  0.816  & *0.970  &   0,\,22~\, \\ 
      {\sc Kov} & $\Lambda_1$    &  0.677  & 0.727  & 0.838 &  *0.953  &  2,\,19~\, \\ 
      {\sc Kow} & $\Lambda_2$      & 0.953    & 0.952 & 0.964 &  *0.967   &  0,\,55~\, \\ 
      {\sc Kow} & $\Lambda_8$    & 0.927    & 0.937   & *0.952 & 0.950  &  0,\,64~\, \\ 
      {\sc Lp} & $\Lambda_2$    & 0.856    &  0.867  & 0.844 &   *0.928 &   0,\,89~\, \\
      {\sc Lp} & $\Lambda_8$    & 0.840  &  0.859   & 0.807 & *0.914  & 0,\,109 \\ 
      {\sc Mp} & $\Lambda_2$   & 0.810   & 0.800 & 0.831  & *0.873   & 0,\,51~\,   \\ 
      {\sc Mp} & $\Lambda_8$   & 0.810   & 0.820  & 0.807 & *0.898   &  0,\,58~\, \\ 
      {\sc OptR} &   $\Lambda_2$  & 0.825  & 0.918   & 0.876 &  *0.970 &   0,\,85~\, \\ 
      {\sc OptR} &   $\Lambda_4$     & 0.825   &  0.878   & 0.870 & *0.970 &  0,\,69~\, \\ 
      {\sc RfIdT} &  $\Lambda_{10}$    & 0.000  & 0.453 & 0.425 & *0.775  &  0,\,43~\, \\    
      {\sc Sl} & $\Lambda_2$  & 0.808   &  0.848  & 0.784  & *0.894  & 0,\,82~\,   \\ 
      {\sc Sl} & $\Lambda_8$     & 0.808  & 0.861  & 0.828  &  *0.897   &  0,\,74~\, \\ 
      {\sc SfT} & $\Lambda_3$     & 0.927 & 0.859     & 0.847 &  *0.941  & 0,\,36~\,  \\ 
      {\sc Vis} &  $\Lambda_3$    & 0.893  & 0.929   & 0.911  & *0.973 &  0,\,43~\, \\ 
      {\sc Homo} &  $\Lambda_9$   & *0.841   & 0.689 & 0.689   & 0.804 &  0,\,87~\,  \\   
      {\sc Lumo} &  $\Lambda_9$   & 0.841 & 0.860  & 0.833 & *0.920  & 0,\,102  \\ 
      {\sc Gap} & $\Lambda_9$   & 0.784  & 0.795 &  0.755 &  *0.876  & 0,\,83~\, \\ 
      {\sc Alpha} & $\Lambda_9$   & 0.961 & 0.888   & 0.953 & *0.980 &  0,\,104  \\ 
      {\sc Cv} &  $\Lambda_9$    & 0.970 & 0.911    & 0.966  &  *0.978  &  0,\,83~\, \\ 
      {\sc mu} & $\Lambda_9$   & 0.367 &  0.409  & 0.403 &  *0.645 &   0,\,112 \\ 
      \bottomrule
  \end{tabular}\end{center}\label{table:phase1b2}
\end{table}


\begin{table}[h!]\caption{Results of constructing prediction functions for polymers.} 
  \begin{center}
    \begin{tabular}{@{} c l     r  r r r   r   c c c c c c @{}}\toprule
      $\pi$ & ~$\Lambda$  &       LLR & ANN  &    ALR    & R-MLR  
        & $K^*_{\mathrm{1}},K^*_{\mathrm{2}}$  \\ \midrule
      {\sc AmD} & $\Lambda_2$   &  0.914    & 0.885   & *0.933 & 0.906  &   0,\,5~\, \\
      {\sc AmD} &  $\Lambda_{13}$  & 0.918     &  0.824  & 0.917 & *0.953   & 0,\,6~\,  \\
      {\sc ChaR} &  $\Lambda_2$  & 0.210  &0.642 & 0.863 & *0.938 &   0,\,10 \\
      {\sc ChaR} &  $\Lambda_{12}$  & 0.088  &  0.640  &  0.835  &  *0.924  & 0,\,9~\,      \\
      {\sc ChaR}    &  $\Lambda_6$  &-0.073  &  0.527 &   0.766 &   *0.950 &   0,\,12  \\ 
      {\sc DeiC} &  $\Lambda_{12}$  & 0.761  &  0.641     & 0.918 & *0.956  &  3,\,41  \\ 
      {\sc DisF} &   $\Lambda_{13}$ &  0.623   & 0.801   & 0.308   & *0.906  & 1,\,23 \\
      {\sc Prm} & $\Lambda_2$   & 0.801  &  0.801 & 0.505 & *0.967  &  0,\,26 \\
      {\sc Prm} &  $\Lambda_{13}$     &0.784    & 0.735  & 0.489 &  *0.977  &  0,\,34  \\
      {\sc RfIdP} &  $\Lambda_{11}$ & 0.104  & 0.423 &  0.853  & *0.962 &   2,\,52  \\
      {\sc RfIdP} &  $\Lambda_{14}$ & 0.373  &0.560  &0.848  & *0.953 & 2,\,43 \\
      {\sc RfIdP}    & $\Lambda_{15}$ & 0.346  & 0.492 &  0.883 &  *0.947 &   5,\,53  \\%
      {\sc Tg} & $\Lambda_2$   & 0.902  &  0.883   & 0.923  & *0.958  & 1,\,33 \\
      {\sc Tg} & $\Lambda_7$   & 0.894  & 0.860   & 0.927 &  *0.957  &  0,\,32  \\
      \hline
  \end{tabular}\end{center}\label{table:phase1b3}
\end{table}

There are 47 instances for constructing prediction functions in 
 Tables~\ref{table:phase1b2} and \ref{table:phase1b3}. 
From these tables, we observe that
method  (iv)  using quadratic descriptors performs better than
methods (i)-(iii) with linear descriptors only 
in 43 out of the 47 instances. 
The averages of the median test R$^2$ of the method (i)-(iv) over  the 47 instances
  are 
 $0.634$, $0.733$, $0.764$ and $0.913$, respectively.
In particular, method  (iv) considerably  improved the performance
for $\pi\in \{${\sc Bp}, {\sc Cp}, {\sc Dc}, {\sc ET},  {\sc Fp}, {\sc Hv},  {\sc Kov}, {\sc Lp},
{\sc RfIdT}, {\sc Gap}, {\sc ChaR}, {\sc DisF},  {\sc Prm}, {\sc RfIdP}$\}$. 
We also see that most descriptors in the resulting descriptor set $D^*$ in R-MLR are quadratic.

\subsection{Results on the Second Phase of the Framework}

To execute the second phase, 
we used a set of seven instances
$I_{\mathrm{a}}$, $I_{\mathrm{b}}^i, i\in[1,4]$, $I_{\mathrm{c}}$
 and $I_{\mathrm{d}}$ based on the seed graphs prepared by Zhu et~al.~\cite{ZAHZNA21}. 
We here present their seed graphs $\GC$ 
(see Appendix~\ref{sec:specification} for the details of $I_{\mathrm{a}}$
and Appendix~\ref{sec:test_instances} for the details of 
$I_{\mathrm{b}}^i, i\in[1,4]$, $I_{\mathrm{c}}$  and $I_{\mathrm{d}}$).

\begin{figure}[h!] \begin{center}
\includegraphics[width=.85\columnwidth]{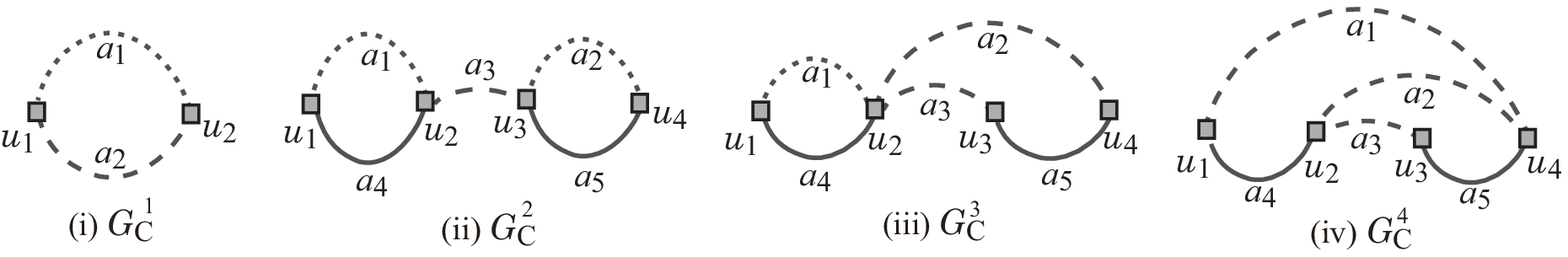}
\end{center} \caption{
(i)  Seed graph   $\GC^1$ for $I_{\mathrm{b}}^1$ and  $I_{\mathrm{d}}$;
(ii) Seed graph $\GC^2$   for $I_{\mathrm{b}}^2$; 
(iii) Seed graph  $\GC^3$   for $I_{\mathrm{b}}^3$; 
(iv)  Seed graph $\GC^4$  for $I_{\mathrm{b}}^4$. }
\label{fig:specification_example_polymer}
\end{figure}

The seed graph  $\GC$ of  $I_{\mathrm{a}}$ is given
 by the graph in Figure~\ref{fig:specification_example_1}(a).
The seed graph $\GC^1$ of  $I_{\mathrm{b}}^1$
(resp., $\GC^i, i=2,3,4$ of $I_{\mathrm{b}}^i,  i=2,3,4$) is illustrated
 in Figure~\ref{fig:specification_example_polymer}.
 
\begin{figure}[!htb]
\begin{center} 
 \includegraphics[width=.69\columnwidth]{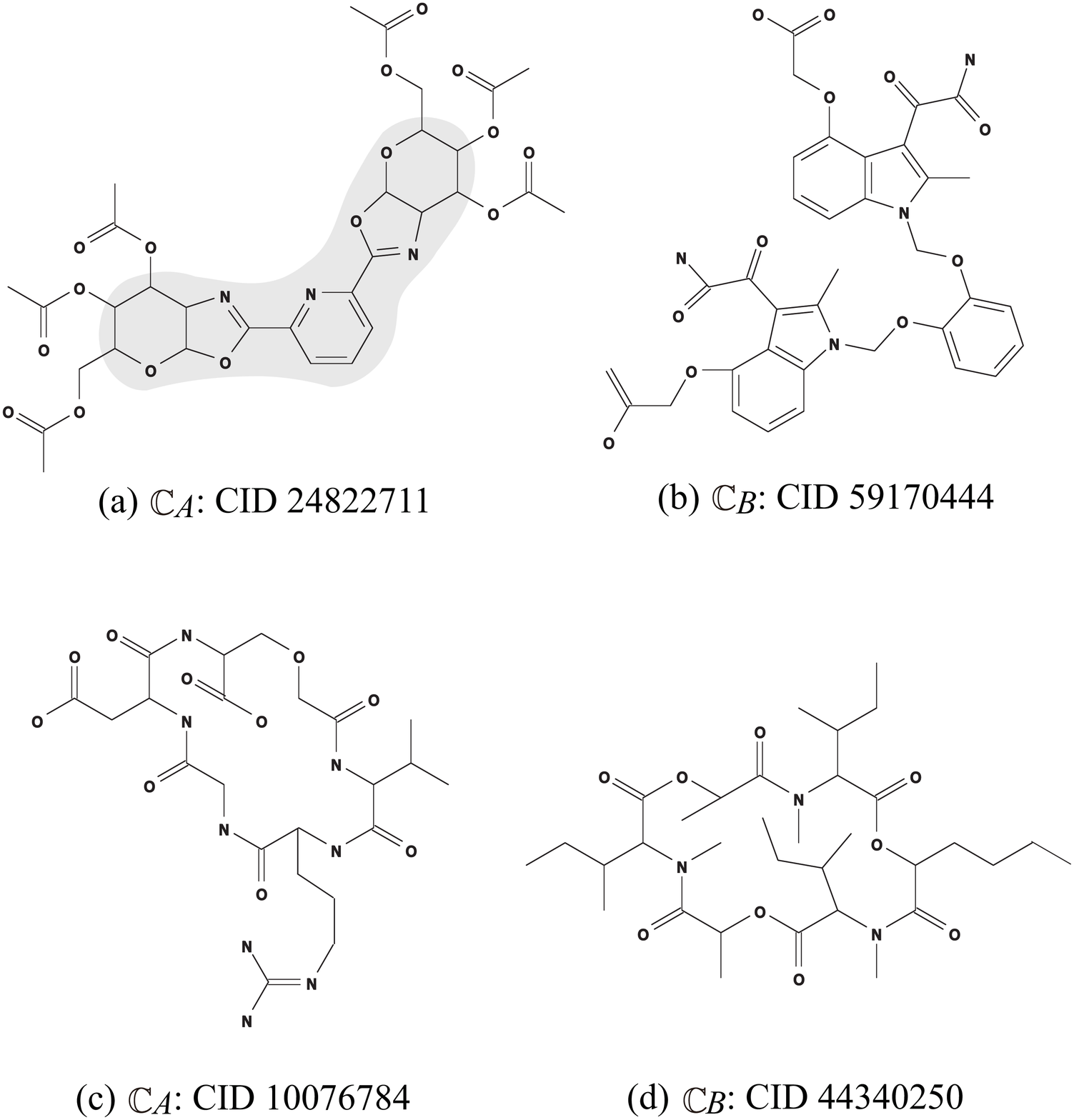}
\end{center}
\caption{An illustration of  chemical compounds 
  for instances  $I_{\rm c}$  and  $I_{\rm d}$: 
(a) $\Co_A$: CID~24822711;
(b)  $\Co_B$: CID~59170444; 
(c) $\Co_A$: CID~10076784;
(d)  $\Co_B$: CID~44340250,
where hydrogens are omitted. 
}
\label{fig:instance_I_c_I_d}  
\end{figure}

Instance  $I_{\mathrm{c}}$ has been introduced 
in order to infer a chemical graph $\Co^\dagger$ such that \\ 
- a core part of $\Co^\dagger$  is equal to that of 
chemical graph $\Co_A$: CID~24822711 in Figure~\ref{fig:instance_I_c_I_d}(a) \\
~ (where the seed graph  $\GC$ of   $I_{\mathrm{c}}$ is indicated 
by the shaded area in Figure~\ref{fig:instance_I_c_I_d}(a)). \\
-  the frequency of each edge-configuration in the non-core of $\Co^\dagger$
is equal to that of chemical graph \\
~   $\Co_B$:  CID~59170444  in  Figure~\ref{fig:instance_I_c_I_d}(b).

Instance  $I_{\mathrm{d}}$ has been introduced 
in order to   infer a chemical  graph $\Co^\dagger$ such that \\
-   $\Co^\dagger$ is monocyclic (where 
the seed graph  of    $I_{\mathrm{d}}$  is given by  $\GC^1$  
 in Figure~\ref{fig:specification_example_polymer}(i)); and  \\
-  the frequency vector of  edge-configurations in  $\Co^\dagger$
is a vector obtained by merging those of \\
~  chemical graphs $\Co_A$: CID~10076784   and $\Co_B$: CID~44340250 
in   Figure~\ref{fig:instance_I_c_I_d}(c) and (d), respectively.   

\medskip \noindent
{\bf Solving an MILP for the inverse problem.  } 
We executed the stage of solving an MILP to infer a chemical graph
 for two properties $\pi\in \{${\sc Bp,  Dc}$\}$.  

 For the MILP formulation  $\mathcal{M}_{f,\eta,\sigma}$,
we use the prediction function  $\eta$ for each $\pi\in \{${\sc Bp,  Dc}$\}$ 
 by method (iv), R-MLR 
 that attained the median  test $\mathrm{R}^2$ in Table~\ref{table:phase1b2}.
 To solve an MILP with the formulation, we used 
{\tt  CPLEX} version 12.10.
Tables~\ref{table:stages_4_5_Bp} and \ref{table:stages_4_5_Dc}  show
   the computational results of the experiment
in this stage for the two properties, 
 where we denote the following:
\begin{enumerate} [nosep,  leftmargin=*]
  
\item[-]  
$n_\LB$: a lower bound on the number of non-hydrogen atoms 
in  a chemical graph $\Co$ to be inferred; 

\item[-]  
  $ \underline{y}^*,~\overline{y}^* $:  
 lower and upper bounds $\underline{y}^*, \overline{y}^*\in \R$ 
  on the value $a(\Co)$ of a chemical graph $\Co$ to be inferred; 
  
\item[-]  
 $\#$v (resp.,  $\#$c): 
 the number  of variables (resp., constraints)  in the MILP;  
  
\item[-]   
 I-time: the   time (sec.) to solve the MILP; 

\item[-]  
    $n$:  the number  $n(\Co^\dagger)$  of  non-hydrogen atoms
     in the chemical graph $\Co^\dagger$   inferred by solving the MILP;   
     
\item[-]  
  $\nint$:  the number  $\nint(\Co^\dagger)$ of interior-vertices in
  the chemical graph $\Co^\dagger$; and 
      
\item[-]  
$\eta$: the predicted property value 
$\eta(f(\Co^\dagger))$ of the chemical graph $\Co^\dagger$.
\end{enumerate}

\begin{table}[h!]\caption{Results of inferring a chemical graph $\Co^\dagger$ 
and generating recombination solutions for  {\sc Bp} with $\Lambda_7$.}  
 \begin{center}
 \begin{tabular}{@{}  c  c   c  r r r r r c  r r r     @{}}\hline                
 inst. & $n_\LB$ &  $ \underline{y}^*,~\overline{y}^* $ & $\#$v~  &  $\#$c~~   &  
   {\small I-time}\!\! & $n$~  &  \!\!$\nint$  &  $\eta $ \!\! & 
                 {\small  D-time} &  {\small $\Co$-LB} &  {\small $\#\Co$}    \\ \hline
   $I_{\mathrm{a}}$ & 30 &  225,\,235  &10502 &10240 & 4.29  & 49 & 26 & 233.92 & 0.072  & 3 & 3       \\%
  $I_{\mathrm{b}}^1$ & 35 &  285,\,295 & 10507 & 7793 & 2.27 & 35 & 10 & 286.52  & 0.034  & 6 & 6   \\%
  $I_{\mathrm{b}}^2$ & 45 &  365,\,375 & 13000 & 10913 & 11.9  & 49 & 25 & 370.70  & 0.14 & 3202 & 100 \\%
  $I_{\mathrm{b}}^3$ & 45 & 305,\,315 & 12788 & 10920 & 7.07  & 48 & 25 & 309.39  & 0.22  & 6304  &100  \\%
  $I_{\mathrm{b}}^4$ & 45 & 260,\,270 & 12576 & 10928 & 10.7 & 49  & 27 & 266.26 & 0.17 & 376 & 100  \\%
  $I_{\mathrm{c}}$   & 50 &  340,\,350 & 7515 & 8270 & 0.867 & 50 & 33 & 344.98 & 0.019  & 2 & 2  \\%
  $I_{\mathrm{d}}$   & 40 &   320,\,330 & 6135 & 7773 & 8.22 & 45 & 23 & 329.85  & 8.3  &6733440 & 100\\
   \hline
   \end{tabular}\end{center}\label{table:stages_4_5_Bp}
\end{table}

 \begin{table}[h!]\caption{ Results of inferring a chemical graph $\Co^\dagger$
 and generating recombination solutions  for {\sc Dc} with $\Lambda_7$.}  
 \begin{center}
 \begin{tabular}{@{}  c  c   c  r r r r r c  r r r     @{}}\hline                
 inst. & $n_\LB$ & $ \underline{y}^*,~\overline{y}^* $ & $\#$v~  &  $\#$c~~   &  
   {\small I-time}\!\! & $n$~  &  \!\!$\nint$  &   $\eta$   & 
                 {\small  D-time} &  {\small $\Co$-LB} &  {\small $\#\Co$}    \\ \hline 
   $I_{\mathrm{a}}$ & 30 &  0.55,\,0.60 &  10194  & 9787  & 3.91  & 41  & 25  & 0.558  & 0.069   & 2  & 2  \\%
  $I_{\mathrm{b}}^1$ & 35 &  1.10,\,1.15  & 10415 &  7368 &  4.73   & 35  & 11  & 1.104  & 0.10  &  16 &  16 \\%
  $I_{\mathrm{b}}^2$ & 45 &  6.00,\,6.05 &  12976  & 10481  & 57.4  & 45 &  25 &  6.04  & 0.12   & 2040  & 100 \\%
  $I_{\mathrm{b}}^3$ & 45 & 1.45,\,1.50  & 2767  & 10488  & 39.7  & 49  & 26 &  1.488  & 0.28 &  21600 &  100 \\%
  $I_{\mathrm{b}}^4$ & 45 & 6.10,\,6.15 & 12558  & 10494  & 26.4  &  46 &  25 &  6.10  & 0.027 &  2  & 2  \\%
  $I_{\mathrm{c}}$   & 50 &  12.35,\,12.40 &  7207 &  7819  & 1.75  & 50 &  34  & 12.38 &  0.020  & 2 &  2  \\%
  $I_{\mathrm{d}}$   & 40 & 3.15,\,3.20  & 5827  & 7325  & 14.9   & 41 &  23 &  3.199 &  0.079  &  18952  & 100 \\
   \hline
   \end{tabular}\end{center}\label{table:stages_4_5_Dc}
\end{table}

\begin{figure}[!htb]
\begin{center} 
\includegraphics[width=.98\columnwidth]{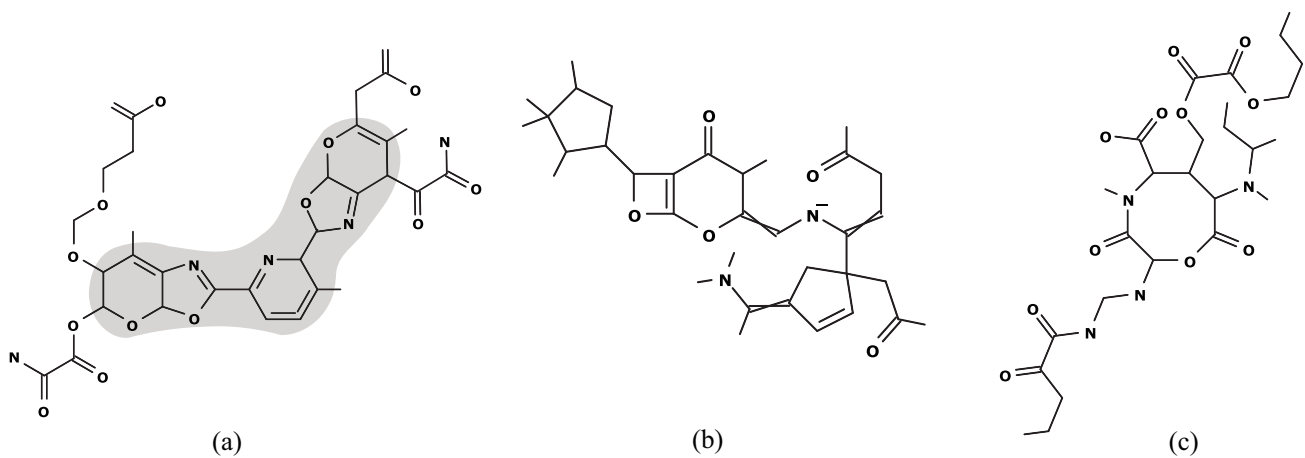}
\end{center}
\caption{ 
(a)  $\Co^\dagger$ with   $\eta(f(\Co^\dagger))=  344.98$ inferred
  from   $I_{\mathrm{c}}$ with $(\underline{y}^*, \overline{y}^*) =(340, 350)$  of  {\sc Bp};   
(b)  $\Co^\dagger$ with   $\eta(f(\Co^\dagger))=0.558$ inferred
 from   $I_{\mathrm{a}}$ with $(\underline{y}^*, \overline{y}^*) =(0.55, 0.60)$  of  {\sc Dc};  and
(c)  $\Co^\dagger$ with   $\eta(f(\Co^\dagger))=3.199$ inferred
 from   $I_{\mathrm{d}}$ with $(\underline{y}^*, \overline{y}^*) =(3.15, 3.20)$  of  {\sc Dc}.   
}
\label{fig:MILP_solutions}  
\end{figure}

 Figure~\ref{fig:MILP_solutions}(a) illustrates  the chemical graph  $\Co^\dagger$  inferred
 from   $I_{\mathrm{c}}$ with $(\underline{y}^*, \overline{y}^*) =(340, 350)$   of  {\sc Bp}
  in Table~\ref{table:stages_4_5_Bp}.  
 
Figure~\ref{fig:MILP_solutions}(b) 
(resp.,  Figure~\ref{fig:MILP_solutions}(c)) illustrates  the chemical graph  $\Co^\dagger$  inferred 
 from  $I_{\mathrm{a}}$ with $(\underline{y}^*, \overline{y}^*) =(0.55, 0.60)$  
 (resp.,  $I_{\mathrm{d}}$ with $(\underline{y}^*, \overline{y}^*) =(3.15, 3.20)$) 
 of  {\sc  Dc}   in Table~\ref{table:stages_4_5_Dc}.

In this experiment, we prepared several different types of instances:
 instances  $I_{\mathrm{a}}$ and $I_{\mathrm{c}}$ have restricted seed graphs,
the other  instances  have abstract seed graphs and
instances $I_{\mathrm{c}}$ and $I_{\mathrm{d}}$ have restricted set of fringe-trees.
From Tables~\ref{table:stages_4_5_Bp} and \ref{table:stages_4_5_Dc},
we observe that an instance with a large number of variables and constraints 
takes more running time than those with a smaller size in general.
All instances in this experiment are solved in a few seconds to around 60 seconds
with our MILP formulation. 


\medskip \noindent
{\bf Generating recombination solutions.  } 
Let   $\Co^\dagger$ be a chemical graph obtained by solving
the MILP $\mathcal{M}_{f,\eta,\sigma}$ for the inverse problem.
We here execute  a stage of generating recombination solutions
 $\C^*\in \mathcal{G}_\sigma$ of $\C^\dagger$
 such that $f(\C^*)=x^*=f(\C^\dagger)$.
 
We execute an  algorithm for generating chemical isomers of   $\Co^\dagger$
up to 100 when the number of all chemical isomers exceeds 100.
For this, we use a dynamic programming algorithm~\cite{ZAHZNA21}. 
The algorithm first decomposes $\Co^\dagger$ into a set of acyclic chemical graphs,
next replaces each acyclic chemical graph $T$ with another  acyclic chemical graph $T'$ that admits
the same feature vector as that of $T$
 and finally assembles the resulting acyclic chemical graphs
into a chemical isomer $\Co^*$ of $\Co^\dagger$. 
The algorithm can compute a lower bound 
on the total number of all chemical isomers $\Co^\dagger$
without generating all of them.

Tables~\ref{table:stages_4_5_Bp} and \ref{table:stages_4_5_Dc}  show
   the computational results of the experiment
in this stage  for the two properties $\pi\in \{${\sc Bp,  Dc}$\}$, 
 where we denote the following:
\begin{enumerate}[nosep,  leftmargin=*]
\item[-]
 D-time: the running time (sec.) to execute the dynamic programming algorithm
  to compute a lower bound on the number 
 of all chemical isomers  $\Co^*$ of  $\Co^\dagger$   
 and generate all (or up to 100) chemical isomers $\Co^*$;
 
\item[-]
 $\Co$-LB: a lower bound on the number of all chemical isomers $\Co^*$ of 
$\C^\dagger$;~and

\item[-]
 $\#\Co$: the number of all (or up to 100) chemical isomers $\Co^*$ of  $\Co^\dagger$  
 generated in this stage.
\end{enumerate} 
  
From Tables~\ref{table:stages_4_5_Bp} and \ref{table:stages_4_5_Dc}, we observe  
 the running time and the number of generated recombination solutions  in this stage. 

The chemical graph $\Co^\dagger$   in  $I_{\mathrm{b}}^2$, $I_{\mathrm{b}}^3$
and $I_{\mathrm{d}}$ admits a large number of 
chemical isomers $\Co^*$ in some cases, 
where a lower bound $\Co$-LB  on the number of chemical isomers
is derived without generating all of them.  
For the other instances, the running time
 for generating up to 100 target chemical graphs in this stage is less than 0.03 second.  
 For some chemical graph $\Co^\dagger$, the number of  chemical isomers found by our algorithm was small.
 This is because some of acyclic chemical graphs in the decomposition of $\Co^\dagger$
 has no alternative acyclic chemical graph other than the original one. 


\medskip \noindent
{\bf Generating neighbor solutions.  } 
Let   $\Co^\dagger$ be a chemical graph obtained by solving
the MILP $\mathcal{M}_{f,\eta,\sigma}$ for the inverse problem.
We executed a stage of generating neighbor solutions 
 of   $\Co^\dagger$.
 
We select an MILP for the inverse problem with a prediction function $\eta$
 such that a solution $\C^\dagger$ of the MILP 
admits only two isomers $\C^*$ in the stage of
generating recombination solutions;
i.e.,  instance 
$I_{\mathrm{c}}$ for property   {\sc Bp} with $\Lambda_7$ and  
 instances  $I_{\mathrm{a}}$,  $I_{\mathrm{b}}^4$ and $I_{\mathrm{c}}$ 
for property   {\sc Dc}  with $\Lambda_7$.

In this experiment, we add to the MILP $\mathcal{M}_{f,\eta,\sigma}$
  an additional set $\Theta$  of two linear constraints 
  on linear and quadratic descriptors as follows.
For the two constraints, we use the prediction functions $\eta_{\pi}$ constructed
by R-MLR for properties  $\pi\in \{${\sc Lp},  {\sc Sl}$\}$ with $\Lambda_8$
in Table~\ref{table:phase1b2}.

Let $D^*_\pi$ denote the set of descriptors selected
in the construction of prediction function for properties
 $\pi\in \{${\sc Bp},{\sc Dc}$\}$ with $\Lambda_7$ and 
 $\pi\in \{${\sc Lp}, {\sc Sl}$\}$ with  $\Lambda_8$ in Table~\ref{table:phase1b2} and
 let  $D^\mathrm{union}_{\pi}, \pi\in \{${\sc Bp}, {\sc Dc}$\}$ denote the union 
  $D^*_\pi \cup D^*_{\mbox{\tiny{\sc Lp}}} \cup D^*_{\mbox{\tiny{\sc Sl}}}$.
We regard each of  $\eta_{\mbox{\tiny{\sc Lp}}}$ and  $\eta_{\mbox{\tiny{\sc Sl}}}$
  as  a function from  $\R^{|D^\mathrm{union}_{\pi}|}$ to $\R$ for $\pi\in\{${\sc Bp}, {\sc Dc}$\}$.
We set $p_{\mathrm{dim}}:=2$ and let  $\Theta$ consist of two 
 linear constraints  $\theta_1:=\eta_{\mbox{\tiny{\sc Lp}}}$
  and $\theta_2:=\eta_{\mbox{\tiny{\sc Sl}}}$.
We set   $\delta:= 0.1$ or 0.05 which defines a two-dimensional grid space 
where  $\C^\dagger$ is mapped to the origin
(see \cite{gridAZHZNA21} for the detail on the neighbors). 
We choose a set $N_0$ of 48 neighbors 
of the origin  $\C^\dagger$ in the grid search space. 
%
For each instance, we check the feasibility of neighbors in $N_0$.
 in a non-decreasing order of the distance between
 the neighbor and the origin.
For each feasible neighbor $z \in N_0$, output a feasible solution $\C^\dagger_z$
of  the augmented MILP instance. 
We set a time limit for checking the feasibility of a neighbor to be 300 seconds, 
and we skip a neighbor when the corresponding MILP is not solved within the time limit.
We also ignore any neighbor $z \in N_0$
without testing the feasibility of $z$ if we find an infeasible neighbor $z'\in N_0$
such that $z'$ is closer to the origin than $z$ is.

Table~\ref{table:neighbor}    shows
   the computational results of the experiment for the three instances, 
 where we denote the following:
\begin{enumerate}[nosep,  leftmargin=*]
\item[-]
(inst., $\pi$):   topological specification $I$ and property $\pi$;

\item[-] $n$: the number of non-hydrogen atoms in the tested instance;  

\item[-] $\delta$: the size of a sub-region in the grid search space;

\item[-] 
\#sol:   the number of new chemical graphs
obtained from the neighbor set $N_0$; 

\item[-] 
\#infs: the number of   neighbors in $N_0$
that are found to be infeasible during the search procedure; 

\item[-] 
\#ign:  the number of   neighbors  in $N_0$
that are ignored during the search procedure; 

\item[-] 
\#TO: the number of  neighbors in $N_0$
such that the time for feasibility check exceeds
the time limit of 300 seconds during the search procedure.

\end{enumerate}

 \begin{table}[h!]\caption{Results of generating neighbor solutions of $\C^\dagger$.  } 
 \begin{center}
 \begin{tabular}{@{}  c   r r r    r r r      @{}}\hline             
(inst., $\pi$) & $n$ & $\delta$~~ &  \#sol  &  \#infs & \#ign  & \#TO   \\ \hline 
   ($I_{\mathrm{c}}$,{\sc Bp})   &  50 & 0.1 & 5  & 1 & 3 & 39     \\ 
  ($I_{\mathrm{a}}$,{\sc Dc})    & 30  &0.1 & 40   & 1 & 0 & 7   \\ 
  ($I_{\mathrm{b}}^4$,{\sc Dc}) & 45  &0.1 & 2    & 0  & 0 & 46     \\ 
  ($I_{\mathrm{c}}$,{\sc Dc})   &  40 &0.05 & 0  & 0 & 0 & 48     \\  
   \hline
   \end{tabular}\end{center}\label{table:neighbor}  
\end{table}
 
The branch-and-bound method for solving an MILP sometimes takes
an extremely large execution time for the same size of instances. 
We introduce a time limit to bound an entire running time to skip such instances
during an execution of testing the feasibility of neighbors in   $N_0$.
From Table~\ref{table:neighbor}, we observe
 that   some number of neighbor solutions of the solution $\C^\dagger$
to  the MILP $\mathcal{M}_{f,\eta,\sigma}$ could be generated for each of the four instances.


\section{Concluding Remarks}\label{sec:conclude}
 
 In the framework of inferring chemical graphs,
 the descriptors of a prediction function were mainly defined
to be the frequencies of local graph structures 
 in the two-layered model and such definition was important to derive
 a compact MILP formulation for inferring a desired chemical graph.
To improve the performance of prediction functions in the framework,
this paper  introduced a multiple of two of these descriptors as a new descriptor
and examined the effectiveness of the new set of descriptors.
For this, we designed a method for reducing the size of a descriptor set
not to lose the learning performance in constructing prediction functions
and gave a compact formulation to compute a product of two values in an MILP.
From the results of our computational experiments, 
we observe that  a  prediction function
constructed by our new method performs considerably better than 
the previous prediction functions for many chemical properties.
We also found that  the modified MILP 
in the second phase of the framework still 
can infer a chemical graph with 
around 50 non-hydrogen atoms. 


\clearpage
 \appendix
\centerline{\bf\LARGE Appendix}

\section{A Full Description of Descriptors}\label{sec:descriptor}

Associated with the two functions 
$\alpha$ and $\beta$ in a chemical graph $\Co=(H,\alpha,\beta)$,
we introduce   functions  
 $\ac: V(E)\to (\Lambda\setminus\{\ttH\})\times (\Lambda\setminus\{\ttH\})\times [1,3]$, 
 $\cs: V(E)\to (\Lambda\setminus\{\ttH\})\times [1,6]$ and
$\ec: V(E)\to ((\Lambda\setminus\{\ttH\})\times [1,6])\times ((\Lambda\setminus\{\ttH\})\times [1,6])\times [1,3]$
in the following. 

 To represent  a feature of the exterior  of  $\Co$, 
  a  chemical rooted tree in $\mathcal{T}(\Co)$ is
  called a {\em fringe-configuration} of $\Co$. 

We also represent leaf-edges in the exterior of $\Co$.
For a leaf-edge $uv\in E(\anC)$ with $\deg_{\anC}(u)=1$, we define
the {\em adjacency-configuration} of $e$ to be an ordered tuple
$(\alpha(u),\alpha(v),\beta(uv))$. 
Define 
\[ \Gac^\lf\triangleq \{(\ta,\tb,m)\mid \ta,\tb\in\Lambda, 
m\in[1,\min\{\val(\ta),\val(\tb)\}]\} \]
as a set of possible adjacency-configurations for leaf-edges. 

To  represent a feature of an interior-vertex $v\in V^\inte(\Co)$ such that
$\alpha(v)=\ta$  and  $\deg_{\anC}(v)=d$
(i.e., the number of non-hydrogen atoms adjacent to $v$ is $d$) 
   in a chemical   graph  $\Co=(H,\alpha,\beta)$,
 we use  a pair $(\ta, d)\in (\Lambda\setminus\{{\tt H}\})\times [1,4]$,
 which we call the {\em chemical symbol} $\cs(v)$ of the vertex $v$.
 We treat $(\ta, d)$ as a single symbol $\ta d$,  and  
define $\Ldg$   to be  the set of all chemical symbols
$\mu=\ta d\in  (\Lambda\setminus\{{\tt H}\})\times [1,4]$.  

We define a method for featuring interior-edges  as follows.
Let $e=uv\in E^\inte(\Co)$  be 
 an interior-edge $e=uv\in E^\inte(\Co)$ 
 such that $\alpha(u)=\ta$, $\alpha(v)=\tb$ and $\beta(e)=m$ 
   in a chemical graph  $\Co=(H,\alpha,\beta)$.
To feature this edge $e$, 
 we use a tuple $(\ta,\tb,m)\in (\Lambda\setminus\{{\tt H}\})
    \times (\Lambda\setminus\{{\tt H}\})\times [1,3]$,
 which we call the {\em adjacency-configuration} $\ac(e)$ of the edge $e$. 
 We introduce a total order $<$ over the elements in $\Lambda$
 to distinguish  between $(\ta,\tb, m)$ and $(\tb,\ta, m)$ 
 $(\ta\neq \tb)$ notationally.
 For a tuple  $\nu=(\ta,\tb, m)$,
 let $\overline{\nu}$ denote the tuple $(\tb,\ta, m)$.

Let $e=uv\in E^\inte(\Co)$  be 
an  interior-edge $e=uv\in E^\inte(\Co)$ 
 such that $\cs(u)=\mu$, $\cs(v)=\mu'$ and $\beta(e)=m$ 
   in a chemical  graph  $\Co=(H,\alpha,\beta)$.
To feature this edge $e$, 
 we use a tuple $(\mu,\mu',m)\in \Ldg\times \Ldg\times [1,3]$, 
 which we call  the {\em edge-configuration} $\ec(e)$ of the edge $e$. 
 We introduce a total order $<$ over the elements in $\Ldg$
 to distinguish between $(\mu,\mu', m)$ and $(\mu', \mu, m)$ 
 $(\mu \neq \mu')$ notationally. 
 For a tuple  $\gamma=(\mu,\mu', m)$,
 let $\overline{\gamma}$ denote the tuple $(\mu', \mu, m)$. 

Let $\pi$ be a chemical property for which we will construct
a prediction function $\eta$ from a feature
vector $f(\Co)$ of a chemical graph $\Co$ 
to a predicted value $y\in \mathbb{R}$
for the  chemical property of $\Co$.

We first choose a set $\Lambda$ of chemical elements
 and then collect a data set  $\mathcal{C}_{\pi}$ of
  chemical compounds  $C$ whose 
  chemical elements belong to $\Lambda$,
  where we regard  $\mathcal{C}_{\pi}$ as a set of chemical graphs $\Co$
  that represent the chemical compounds $C$  in  $\mathcal{C}_{\pi}$.
To define the interior/exterior of 
chemical graphs  $\Co\in \mathcal{C}_{\pi}$,
we  next choose a branch-parameter ${\rho}$, where
 we recommend ${\rho}=2$.  
 
Let $\Lambda^\inte(\mathcal{C}_\pi)\subseteq \Lambda$ 
(resp., $\Lambda^\ex(\mathcal{C}_\pi)\subseteq \Lambda$)
denote the set  of chemical elements  used in
the set $V^\inte(\Co)$ of interior-vertices
(resp., the set $V^\ex(\Co)$ of  exterior-vertices) of $\Co$
 over all chemical graphs $\Co\in \mathcal{C}_\pi$, 
and $\Gamma^\inte(\mathcal{C}_\pi)$
denote the set of edge-configurations used in
the set $E^\inte(\Co)$  of interior-edges in $\Co$
 over all chemical graphs $\Co\in \mathcal{C}_\pi$. 
Let $\mathcal{F}(\mathcal{C}_\pi)$ denote the set of
chemical rooted trees $\psi$  
r-isomorphic to a chemical rooted tree in $\mathcal{T}(\Co)$
  over all chemical graphs $\Co\in \mathcal{C}_\pi$,
  where possibly a chemical rooted tree $\psi\in \mathcal{F}(\mathcal{C}_\pi)$
  consists of a single chemical element $\ta\in \Lambda\setminus \{{\tt H}\}$.
  
We define an integer encoding of a finite set $A$ of elements
to be a bijection $\sigma: A \to [1, |A|]$, 
where we denote by $[A]$   the set $[1, |A|]$ of integers.
Introduce  an integer coding of each of the   sets 
$\Lambda^\inte(\mathcal{C}_\pi)$, $\Lambda^\ex(\mathcal{C}_\pi)$, 
$\Gamma^\inte(\mathcal{C}_\pi)$ and $\mathcal{F}(\mathcal{C}_\pi)$. 
Let $[\ta]^\inte$  
(resp., $[\ta]^\ex$)  denote   
the coded integer of  an element $\ta\in \Lambda^\inte(\mathcal{C}_\pi)$
(resp., $\ta\in \Lambda^\ex(\mathcal{C}_\pi)$),  
$[\gamma]$   denote  
the coded integer of  an element $\gamma$ in $\Gamma^\inte(\mathcal{C}_\pi)$
and 
$[\psi]$   denote  an element $\psi$ in $\mathcal{F}(\mathcal{C}_\pi)$. 
 
 Over 99\% of  chemical compounds $\Co$ with up to
  100 non-hydrogen atoms in  PubChem  have degree at most 4
 in the hydrogen-suppressed graph $\anC$~\cite{AZSSSZNA20}. 
We assume that a chemical graph $\Co$
 treated in this paper satisfies  $\deg_{\anC}(v)\leq 4$
in the hydrogen-suppressed graph $\anC$.
 
In our model, we  use an integer 
  $\mathrm{mass}^*(\ta)=\lfloor 10\cdot \mathrm{mass}(\ta)\rfloor$, 
 for each $\ta\in \Lambda$.
 
 For a chemical property $\pi$,
 we define a set $D_\pi^{(1)}$ of descriptors  
  of a  chemical graph $\C=(H,\alpha,\beta)\in \mathcal{C}_{\pi}$ 
  to be  the following  
non-negative integers $\dcp_i(\C)$, $i\in [1,K_1]$, where 
$K_1= 14+ |\Lambda^\inte(\mathcal{C}_\pi)|+|\Lambda^\ex(\mathcal{C}_\pi)|
         +|\Gamma^\inte(\mathcal{C}_\pi)|+|\mathcal{F}(\mathcal{C}_\pi)|+|\Gac^\lf|$. 


\begin{enumerate}  
\item   
$\dcp_1(\C)$: the number  $|V(H)|-|\VH|$ of non-hydrogen atoms  in  $\C$.  
 
\item   
$\dcp_2(\C)$: the rank 
of   $\C$ (i.e., the minimum number of edges to be removed to make the graph 
acyclic).  

\item 
$\dcp_3(\C)$:  the number $|V^\inte(\C)|$ of interior-vertices in  $\C$.
  
\item 
$\dcp_4(\C)$: 
the average $\overline{\mathrm{ms}}(\C)$ of mass$^*$ 
over all atoms in $\C$; \\
 i.e., $\overline{\mathrm{ms}}(\C)\triangleq 
 \frac{1}{|V(H)|}\sum_{v\in V(H)}\mathrm{mass}^*(\alpha(v))$. 

\item 
$\dcp_i(\C)$,  $i=4+d,   d\in [1,4]$: 
the number $\dg_d^{\oH} (\C)$ 
 of non-hydrogen vertices $v\in V(H)\setminus \VH$
 of degree $\deg_{\anC}(v)=d$
 in the hydrogen-suppressed chemical graph $\anC$.  
 
\item 
$\dcp_i(\C)$,  $i=8+d,   d\in [1,4]$: 
the number $\dg_d^\inte(\C)$
 of interior-vertices of interior-degree  $\deg_{\C^\inte}(v)=d$
  in the interior $\C^\inte=(V^\inte(\C),E^\inte(\C))$ of  $\C$. 
  
   
\item $\dcp_i(\C)$, $i=12+m$,  $m\in[2,3]$: 
the number $\bd_m^\inte(\C)$
 of  interior-edges with bond multiplicity $m$ in  $\C$; 
 i.e., $\bd_m^\inte(\C)\triangleq |\{e\in E^\inte(\C)\mid \beta(e)=m\}|$.

\item $\dcp_i(\C)$, $i=14+[\ta]^\inte$, 
 $\ta\in \Lambda^\inte(\mathcal{C}_\pi)$: 
 the frequency $\na_\ta^\inte(\C)=|V_\ta(\C)\cap V^\inte(\C) |$ 
 of chemical element $\ta$ in
 the set $V^\inte(\C)$ of  interior-vertices in  $\C$. 
 
\item $\dcp_i(\C)$, 
$i=14+|\Lambda^\inte(\mathcal{C}_\pi)|+[\ta]^\ex$, 
 $\ta\in \Lambda^\ex(\mathcal{C}_\pi)$: 
 the frequency $\na_\ta^\ex(\C)=|V_\ta(\C)\cap V^\ex(\C) |$
  of chemical element $\ta$ in
 the set $V^\ex(\C)$ of  exterior-vertices in  $\C$. 
 
\item $\dcp_i(\C)$, 
$i=14+|\Lambda^\inte(\mathcal{C}_\pi)|+|\Lambda^\ex(\mathcal{C}_\pi)|+ [\gamma]$, 
$\gamma \in \Gamma^\inte(\mathcal{C}_\pi)$: 
the frequency $\ec_{\gamma} (\Co)$ of edge-configuration $\gamma$
in the set $E^\inte(\C)$ of interior-edges   in  $\C$.

\item $\dcp_i(\C)$, 
$i= 14+|\Lambda^\inte(\mathcal{C}_\pi)|+|\Lambda^\ex(\mathcal{C}_\pi)|
+ |\Gamma^\inte(\mathcal{C}_\pi)|+ [\psi]$,  
 $\psi \in \mathcal{F}(\mathcal{C}_\pi)$: 
the frequency $\fc_{\psi}(\C)$ of fringe-configuration $\psi $
in the set of ${\rho}$-fringe-trees in  $\C$. 

\item $\dcp_i(\C)$, 
$i= 14+|\Lambda^\inte(\mathcal{C}_\pi)|+|\Lambda^\ex(\mathcal{C}_\pi)|
+ |\Gamma^\inte(\mathcal{C}_\pi)|+|\mathcal{F}(\mathcal{C}_\pi)|+ [\nu]$,  
 $\nu \in \Gac^\lf$: 
the frequency $\ac_{\nu}^\lf(\C)$ of adjacency-configuration $\nu$
in the set of leaf-edges in  $\anC$. 
\end{enumerate} 

In the framework for polymers~\cite{poly_ICZAHZNA22},
a polymer is treated as a chemical graph of its repeating unit,
where we call an edge $e$ a {\em link-edge} when it lays on any path
between the two joint-points of the repeating unit,
and call the end-vertices of a link-edge {\em connecting-vertices}.
The set of descriptors for a polymer is defined analogously with
the above set for a monomer except for 
$\dcp_2(\C)$ is replaced with the number of link-edges
and the following two kinds of descriptors are added:
the frequency $\ec_{\gamma} (\C)$ of edge-configuration $\gamma$
 of link-edges; and the frequency of chemical symbols  
 of connecting-vertices (see \cite{poly_ICZAHZNA22} for the details).

In this paper, we introduce a set of quadratic descriptors.
For this, we first normalize each descriptor  $\dcp_i(\C), i\in[1,K_1]$
to a value $x(i)$ between 0 and 1 by scaling the minimum and   maximum
values to 0 and 1, respectively.
Then construct a set
 $D_\pi^{(2)}:= 
             \{x(i)x(j) \mid 1\leq i\leq j\leq K_1\}
    \cup \{x(i)(1-x(j))\mid 1\leq i, j\leq K_1\}$ of  
 $(3(K_1)^2+K_1)/2$ quadratic descriptors.
 In this paper, we reduce the union $D_\pi^{(1)}\cup D_\pi^{(2)}$
 to a subset to construct a prediction function.


\section{Methods for Reducing Descriptors}\label{sec:reduce_descriptors}

Let $\mathcal{C}$ be a set of components, $D$ be a set of all descriptors
and  $K^*\in [1,|D|]$ be a number of descriptors we want to choose from $D$.

Given a data set $\mathcal{C}$, a set $D$ of descriptors and a real $\lambda>0$, 
let Des-set-LLR$(\mathcal{C},D,\lambda)$ denote 
the set $S$ of descriptors $d\in D$ such that $\w(d)=0$  
for the hyperplane $(\w,b)$ output by LLR$(\mathcal{C},D,\lambda)$
 (where we numerically treat
 $\w(d)$ with $|\w(d)|\leq 10^{-6}$ as 0 in our experiment).

\subsection{A method based on Lasso linear regression}

Since the Lasso linear regression finds some number of descriptors
$d\in D$ with $w(d)=0$ in the output hyperplane $(\w,b)$,
we can reduce a given set of descriptors by applying the Lasso linear regression
repeatedly. 
Choose parameters $c_{\max}$ and $d_{\max}$ so that  LLR$(\mathcal{C},D,\lambda)$
can be executed in a reasonable running time when  $|\mathcal{C}|\leq c_{\max}$
and $|D|\leq d_{\max}$.
Let $\widetilde{K}\in [1,|D|]$ be an integer for the number of descriptors that
we choose from a given set $D$ of descriptors.

\bigskip
\noindent
LLR-Reduce$(\mathcal{C},D)$: \\
Input: A data set $\mathcal{C}$ and a set $D$ of descriptors;\\
Output: A subset $\widetilde{D}\subseteq D$
 with $|\widetilde{D}|=\widetilde{K}$.\\
Initialize $D':=D$; \\
{\bf while} $|D'|>\widetilde{K}$ {\bf do}\\
~~ Partition $D'$ randomly into disjoint subsets $D_1,D_2,\ldots,D_p$ 
    such that $|D_i|\leq d_{\max}$ for each $i$;\\
~~ {\bf for}  each  $i=1,2,...,p$ {\bf do} \\
~~~~ Choose a subset $\mathcal{C}_i$ 
    with $|\mathcal{C}_i|=\min\{c_{\max} ,|\mathcal{C}|\}$ of $\mathcal{C}$ randomly; \\
~~~~ $D'_i:=$Des-set-LLR$(\mathcal{C}_i,D_i,\lambda)$
  for some $\lambda>0$  \\
~~ {\bf  endfor}; \\
~~ $D':=D'_1\cup D'_2\cup \cdots \cup D'_p$\\
{\bf endwhile}; \\
Output $\widetilde{D}:=D'$ after adding  to $D'$
extra $\widetilde{K}-|D'|$ descriptors 
from the previous set $D'$ when  $|D'|<\widetilde{K}$
by using the K-best method.\\

In this paper, we set $c_{\max}:=200$, $d_{\max}:=200$
and $\widetilde{K}:=5000$ in our computational experiment.

\subsection{A method based on backward stepwise procedure}

A backward stepwise procedure~\cite{DS66} 
reduces the number of descriptors one by one 
choosing the one removal of which maximizes
the learning performance and outputs a subset with the maximum learning performance
among all subsets during the reduction iteration.

For a subset $S\subseteq D$ and a positive integer $p$,
 let $\mathrm{R}^2_{\mathrm{CV,MLR}}(\mathcal{C},S,p)$  denote
 $\mathrm{R}^2_{\mathrm{CV}}(\mathcal{C},S,p)$  for constructing
a prediction function $\eta_{w,b}$ by MLR$(\mathcal{C},S)$.
We define 
a performance evaluation function $g_p:2^D\to \R$ for an integer $p\geq 1$ such that
$g_p(S)=\mathrm{R}^2_{\mathrm{CV,MLR}}(\mathcal{C},S,p)$.
The backward stepwise procedure with this function $g_p$
 is described as follows. 

\bigskip
\noindent
BS-Reduce$(\mathcal{C},D,p)$: \\
Input: A data set $\mathcal{C}$, a set $D$ of descriptors, 
an integer $p\geq 1$ and  
a performance evaluation function $g_p:2^D\to \R$ defined above; \\ 
Output: A subset $D^*\subseteq D$.\\
Compute $\ell_{\mathrm{best}}:=g_p(D)$; Initialize $D_{\mathrm{best}}:=D' := D$; \\
{\bf while} $D' \neq \emptyset$ {\bf do}\\
~~ Compute $\ell(d):=g_p(D' \setminus \{d\})$ for each descriptor $d\in D' $; \\
~~ Set $d^*\in D' $ to be a descriptor  that maximizes $\ell(d)$ over all  $d\in D' $; \\
~~ Update: $D' :=D' \setminus \{d^*\}$;\\
~~ {\bf if} $\ell(d^*)>\ell_{\mathrm{best}}$ {\bf then} update $D_{\mathrm{best}}:=D' $ and 
       $\ell_{\mathrm{best}}:=\ell(d^*)$ {\bf endif}\\
{\bf endwhile}; \\
Output $D^*:=D_{\mathrm{best}}$.

\bigskip 
Based on the Lasso linear regression and the backward stepwise procedure,
we design the following method for choosing a subset $D^*$ of 
a given set $D$ of descriptors. 
We are given a give set $A$ of 17 real numbers
and a set $B(a)$ of 16 real numbers close to each number $a\in A$.
The method first choose a best parameter $\lambda_{\mathrm{best}}\in A$
to construct a prediction function by LLR
and then choose a subset $D_i\subseteq D$ for each $\lambda_i\in B(\lambda_{\mathrm{best}})$
by the backward stepwise procedure.
The procedure  takes $O(|D|^2)$ iterations which may take a large amount of running time.
We introduce an upper bound $s_{\max}$ on the size of an input descriptor $D$
for the backward stepwise procedure. 
Let $p_1$, $p_2$ and $p_3$ be integer parameters
that control the number of executions of cross-validations to evaluate
the learning performance in the method.

\bigskip
\noindent
Select-Des-set$(\mathcal{C},D)$: \\
Input: A data set $\mathcal{C}$, a set $D$ of descriptors, \\
a set 
$A=\{0, 10^{-6}, 10^{-5}, 10^{-4}, 10^{-3}, 0.01, 0.05,$ 
$ 0.1,	0.5, 0.75, 1, 2, 5, 10, 25, 50, 100\}$, and \\
a set $B(\lambda)$ of 16 real numbers close to  each $\lambda\in A$; \\
Output: A subset $D^*$ of $D$.\\
{\bf for}   each $\lambda\in  A$ {\bf do}\\
~~ Compute $D_\lambda:=$Des-set-LLR$(\mathcal{C},D,\lambda)$ and  
     $\ell_\lambda:= \mathrm{R}^2_{\mathrm{CV,MLR}}(\mathcal{C},D_\lambda,p_1)$ \\
{\bf endfor}; \\ 
   Set $\lambda_{\mathrm{best}}$ to be a $\lambda\in A$ that maximizes $\ell_\lambda$; 
Denote $B(\lambda_{\mathrm{best}})$ by $\{\lambda_1,\lambda_2,\ldots,\lambda_{16}\}$; \\
{\bf for} each $i\in [1,16]$ {\bf do}\\
~~ Compute $D_i:=$ Des-set-LLR$(\mathcal{C},D,\lambda_i)$ and 
    let $(w,b), w\in \R^{|D|}, b\in \R$ be the hyperplane \\
~~  obtained      by  this LLR; \\
~~ {\bf if} $|D_i|\leq s_{\max}$ {\bf then}\\
~~ ~~ $D'_i:=D_i$\\
~~ {\bf else} \\
~~~~ Let $D'_i$ consist of $s_{\max}$ descriptors $d\in D_i$ that have
the $s_{\max}$ largest absolute values $|w(d)|$ \\
~~~~ in the weight sets $\{w(d)\mid d\in D_i\}$
of the hyperplane $(w,b)$ \\
~~ {\bf endif}; \\
~~     $D_i^\dagger:=$BS-Reduce$(\mathcal{C},D'_i, p_2)$; 
 $\ell_i:= \mathrm{R}^2_{\mathrm{CV,MLR}}(\mathcal{C},D_i^\dagger,p_3)$ \\
{\bf endfor}; \\ 
Set $D^*$ to be a set $D_i^\dagger$ that maximizes $\ell_i, i\in [1,16]$.

\bigskip 
 In our computational experiment in this paper,
 we set $p_1:=p_2:=p_3:=5$ and  $s_{\max}:= 150+10^4/(|\mathcal{C}|+200)$.

 
\section{Specifying Target Chemical Graphs}\label{sec:specification} 

Given a prediction function $\eta$ and 
a target value $y^*\in \mathbb{R}$, 
we call a chemical graph $\C^*$ such that $\eta(x^*)=y^*$
for the feature vector $x^*=f(\C^*)$ a {\em target chemical graph}.
This section  presents a set of rules for 
 specifying  topological substructure
  of a target chemical graph in a flexible way in Stage~4.

We first describe how to reduce a chemical graph $\C=(H,\alpha,\beta)$ into
an abstract form based on which our specification rules will be defined.
To illustrate the reduction process,
we use the chemical graph $\C=(H,\alpha,\beta)$
such that $\anC$ is given in Figure~\ref{fig:example_chemical_graph}.
 
\begin{figure}[h!] \begin{center}
\includegraphics[width=.65\columnwidth]{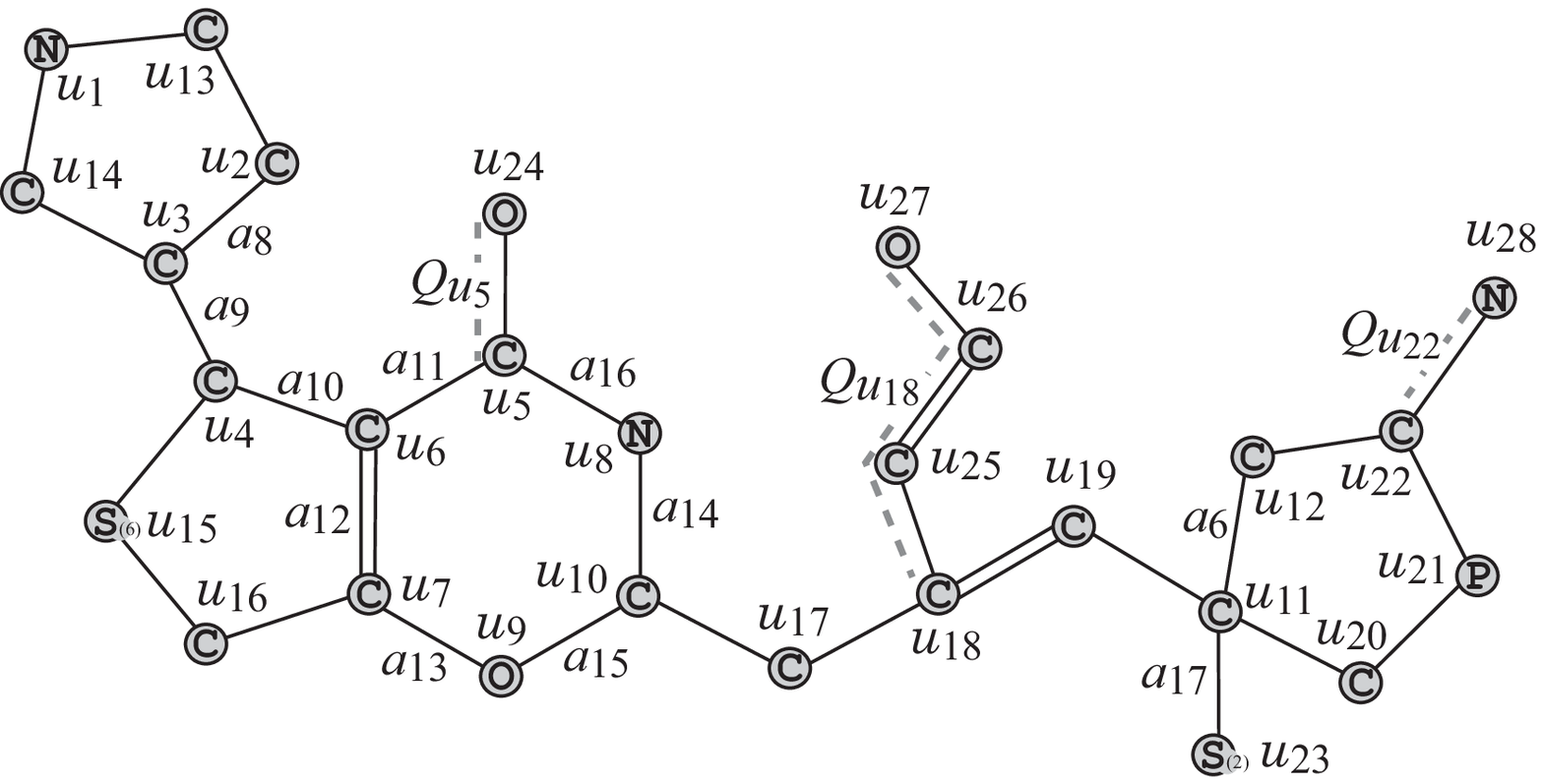}
\end{center} \caption{The interior $H^\inte$ of
chemical graph $\C$ with $\anC$ 
  in Figure~\ref{fig:example_chemical_graph} for ${\rho}=2$.
}
\label{fig:specification_example_interior} \end{figure}

\begin{figure}[h!] \begin{center}
\includegraphics[width=.98\columnwidth]{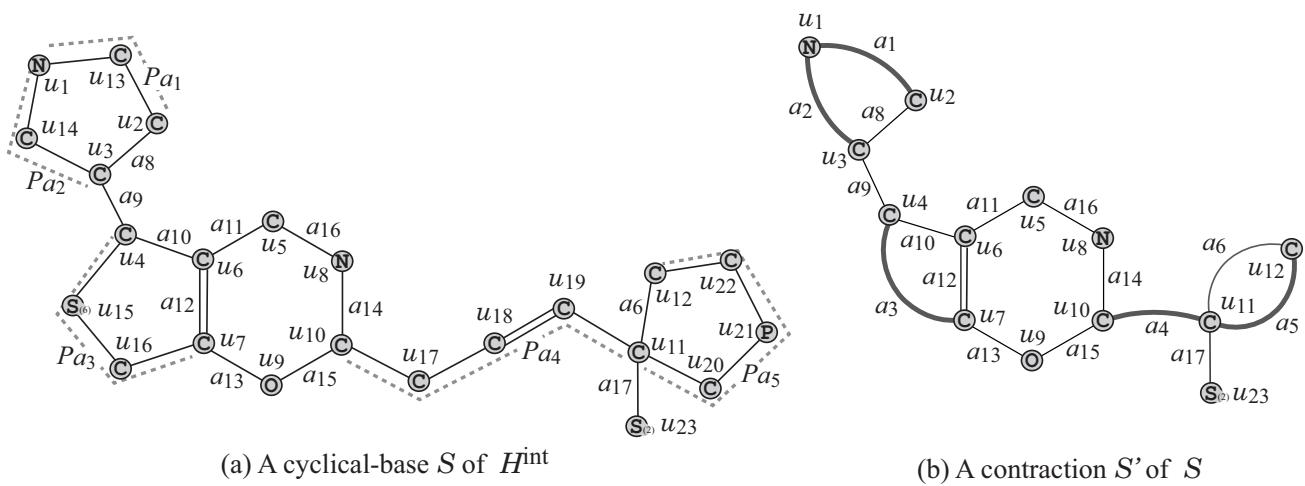}
\end{center} \caption{
(a) A cyclical-base  
$S=H^\inte- \bigcup_{u\in \{u_5,u_{18},u_{22}\}}(V(Q_u)\setminus \{u\})$
of the interior  $H^\inte$ in Figure~\ref{fig:specification_example_interior};
(b) A contraction $S'$ of  $S$ for a pure path set 
 $\mathcal{P}=\{P_{a_1},P_{a_2},\ldots,P_{a_5}\}$ 
in (a),
where a new edge obtained by contracting a pure path is depicted
with a thick line.   
}
\label{fig:specification_example_R2_3} \end{figure}

 \begin{enumerate}
 \item[R1] {\bf Removal of all ${\rho}$-fringe-trees: } 
The interior $H^\inte=(V^\inte(\C),E^\inte(\C))$ of $\C$ 
is obtained by removing the non-root vertices of 
each ${\rho}$-fringe-trees $\C[u]\in\mathcal{T}(\C), u\in V^\inte(\C)$. 
Figure~\ref{fig:specification_example_interior} illustrates
the interior $H^\inte$ of 
chemical graph $\C$ with ${\rho}=2$
  in Figure~\ref{fig:example_chemical_graph}. 
  
 \item[R2] {\bf Removal of some leaf paths: } 
 We call a $u,v$-path $Q$ in $H^\inte$  a {\em leaf path} if 
  vertex $v$ is a leaf-vertex of $H^\inte$
  and the degree of each internal vertex of $Q$  in $H^\inte$  is 2,
  where we regard that $Q$ is rooted at vertex $u$. 
A connected subgraph $S$ of the interior $H^\inte$ of $\C$  
is called a {\em cyclical-base}
if $S$ is obtained from $H$
by removing the vertices in $V(Q_u)\setminus \{u \}, u\in X$ 
for a subset $X$ of interior-vertices  and a set  $\{Q_u \mid u\in X\}$ of leaf 
 $u,v$-paths $Q_u$  such that    
 no two paths $Q_u$ and $Q_{u'}$ share a vertex.
Figure~\ref{fig:specification_example_R2_3}(a) illustrates
a cyclical-base  $S=H^\inte- \bigcup_{u\in X}(V(Q_u)\setminus \{u\})$
of the interior  $H^\inte$  
for a set 
$\{Q_{u_5}=(u_5,u_{24}), 
     Q_{u_{18}}=(u_{18},u_{25},u_{26},u_{27}),
     Q_{u_{22}}=(u_{22},u_{28})\}$ of leaf  paths 
in Figure~\ref{fig:specification_example_interior}.  

 \item[R3] {\bf Contraction of some pure paths: } 
 A path in $S$ is called {\em pure} 
 if  each internal vertex of the path  is of degree 2. 
 Choose a set $\mathcal{P}$ of several pure paths in $S$ 
 so that no two paths share  vertices except for their end-vertices. 
 A graph $S'$ is called a {\em contraction} of a graph $S$
  (with respect to $\mathcal{P}$) 
 if $S'$ is obtained from $S$ by replacing 
 each pure $u,v$-path  with a single edge $a=uv$,
 where $S'$ may contain multiple edges between the same pair of adjacent vertices.
Figure~\ref{fig:specification_example_R2_3}(b) illustrates
a contraction $S'$ obtained from 
the chemical graph  $S$
by contracting each $uv$-path $P_a\in  \mathcal{P}$ into a new edge $a=uv$,
where $a_1=u_1 u_{2},  a_2=u_1 u_{3},  a_3=u_4 u_{7}, a_4=u_{10}u_{11}$
and $a_5=u_{11}u_{12}$ and 
 $\mathcal{P}=\{
 P_{a_1}=(u_1,u_{13},u_{2}), 
 P_{a_2}=(u_{1},u_{14},u_{3}),
 P_{a_3}=(u_{4},u_{15},u_{16},u_{7}), 
 P_{a_4}=(u_{10},u_{17},u_{18},u_{19},u_{11}),
 P_{a_5}=(u_{11},u_{20},u_{21},u_{22},u_{12})\}$ of pure paths 
in Figure~\ref{fig:specification_example_R2_3}(a). 
\end{enumerate}

We will define a set of rules so that 
a chemical graph can be obtained 
from a graph (called a seed graph in the next section)
by applying processes R3 to R1 in a reverse way. 
We specify topological substructures of a target chemical graph
with a tuple  $(\GC,\sint,\sce)$  called  a {\em target specification}
defined under the set of the following rules. 

\subsection*{Seed Graph}

A  {\em seed graph} $\GC=(\VC,\EC)$ is defined
to be a graph (possibly with multiple edges) such that 
the edge set $\EC$ consists of four sets 
$\Et$, $\Ew$, $\Ez$ and $\Eew$, 
where each of them can be empty.
A seed graph plays a role of the most abstract form $S'$ in R3.  
Figure~\ref{fig:specification_example_1}(a) illustrates an example of a seed graph
$\GC$ with $\mathrm{r}(\GC)=5$,   
where $\VC=\{u_1,u_2,\ldots,u_{12},u_{23}\}$, 
$\Et=\{a_1,a_2,\ldots,a_5\}$, 
$\Ew=\{a_6\}$,
$\Ez=\{a_7\}$ and 
$\Eew=\{a_8,a_9,\ldots,a_{16}\}$.

 A {\em subdivision} $S$ of $\GC$  
is a graph constructed from a seed graph $\GC$ 
according to the following rules:
\begin{enumerate}[leftmargin=*]
\item[-]
Each edge $e=uv\in \Et$ is replaced
with a $u,v$-path $P_e$ of length at least 2;

\item[-] 
Each edge $e=uv\in \Ew$ is replaced
with a $u,v$-path $P_e$ of length at least 1
(equivalently $e$ is directly used or replaced with
a $u,v$-path $P_e$ of length at least 2);

\item[-] 
Each edge $e\in \Ez$ is either used or discarded, where 
 $\Ez$ is required to be chosen as a non-separating edge subset of
 $E(\GC)$ since otherwise the connectivity of a final chemical graph $\Co$
 is not guaranteed; 
$\mathrm{r}(\Co)= \mathrm{r}(\GC)-|E'|$ holds
for a subset $E'\subseteq \Ez$ of edges discarded 
in a  final chemical graph $\Co$; 
and 

\item[-]
Each edge $e\in \Eew$ is always used directly. 
\end{enumerate}

We allow a possible elimination of edges in $\Ez$ as an optional rule
in constructing a target chemical graph from a seed graph, 
even though such an operation has 
not been included in the process R3. 
A subdivision  $S$ plays a role of a cyclical-base in R2. 
A target chemical graph $\C=(H,\alpha,\beta)$ will contain  $S$  as a subgraph
of the interior $H^\inte$ of $\C$.


\subsection*{Interior-specification}

A graph $H^*$ that serves as the interior $H^\inte$ of
a target chemical graph $\C$ will be constructed as follows.
First construct a subdivision  $S$ of a seed graph $\GC$ 
by replacing each edge $e=u u'\in \Et\cup\Ew$
with a pure $u,u'$-path $P_e$.
Next construct a supergraph $H^*$ of $S$ by 
attaching a leaf path $Q_v$ at each vertex $v\in \VC$ or
at an internal vertex $v\in V(P_e)\setminus\{u,u'\}$ 
of each pure $u,u'$-path $P_e$ for some edge $e=uu'\in \Et\cup\Ew$,
where possibly $Q_v=(v), E(Q_v)=\emptyset$ 
(i.e., we do not attach any new edges to $v$).
We introduce the following rules for specifying
 the size of $H^*$, the length $|E(P_e)|$  of
a pure path  $P_e$,  the length $|E(Q_v)|$ of
a   leaf path $Q_v$, the number of  leaf paths $Q_v$
and a bond-multiplicity of each interior-edge,
where we call the set of prescribed constants  
 an  {\em interior-specification}   $\sint$: 
\begin{enumerate}[leftmargin=*]
 \item[-]
  Lower and upper bounds $\nint_\LB, \nint_\UB\in \mathbb{Z}_+$ 
  on   the number of interior-vertices 
of a target chemical graph~$\C$. 
  
\item[-] 
For each edge $e=u u'\in \Et\cup\Ew$, 
\begin{description} 
\item[]
 a lower bound $\ell_{\LB}(e)$ and 
 an upper bound $\ell_{\UB}(e)$  on the length $|E(P_e)|$ of
 a pure $u,u'$-path $P_e$. 
(For a notational convenience, set 
$\ell_\LB(e):=0$, $\ell_\UB(e):=1$, $e\in \Ez$ and
$\ell_\LB(e):=1$, $\ell_\UB(e):=1$, $e\in \Eew$.)
   
\item[]  
 a lower bound $\bl_{\LB}(e)$ and 
 an upper bound $\bl_{\UB}(e)$ on the number of leaf paths $Q_v$ attached 
 at  internal vertices $v$ of a pure $u,u'$-path $P_e$.   

\item[] 
 a lower bound $\ch_{\LB}(e)$ and 
 an upper bound $\ch_{\UB}(e)$  on the maximum 
 length  $|E(Q_v)|$ of a leaf path $Q_v$ attached  
 at an internal vertex $v\in V(P_e)\setminus\{u,u'\}$ 
 of a pure $u,u'$-path $P_e$.   
\end{description} 

\item[-]
For each vertex $v\in \VC$, 
\begin{description} 
\item[]
 a lower bound $\ch_{\LB}(v)$ and 
 an upper bound $\ch_{\UB}(v)$  on  
 the number of leaf paths $Q_v$ attached to $v$,
 where $0\leq \ch_{\LB}(v)\leq \ch_{\UB}(v)\leq 1$.
 
\item[]
 a lower bound $\ch_{\LB}(v)$ and 
 an upper bound $\ch_{\UB}(v)$  on the
 length $|E(Q_v)|$ of a leaf path $Q_v$ attached to $v$. 
\end{description}  

\item[-] 
For each edge $e=u u'\in \EC$, 
a lower bound $\bd_{m, \LB}(e)$ 
and an  upper bound $\bd_{m, \UB}(e)$  on
the number of edges with bond-multiplicity $m\in [2,3]$ in
$u,u'$-path $P_e$, where we regard $P_e$, $e  \in \Ez\cup \Eew$ 
as single edge $e$.
\end{enumerate}

We call a graph $H^*$ that satisfies an interior-specification $\sint$
a {\em $\sint$-extension of $\GC$}, 
where the bond-multiplicity of each edge has been determined.

Table~\ref{table:interior-spec}  shows an example of 
an interior-specification  $\sint$ to the seed graph  $\GC$ in 
Figure~\ref{fig:specification_example_1}. 

\begin{table}[h!]\caption{Example~1 of an interior-specification  $\sint$. }
\begin{tabular}{ |  c | c |  } \hline 
$\nint_\LB=20$ & $\nint_\UB = 28$ \\\hline 
\end{tabular}

 \begin{tabular}{ |  c | c c c c c c |  } \hline
                        & $a_1$ &  $a_2$ &   $a_3$ &   $a_4$ &   $a_5$ &   $a_6$   \\\hline
 $\ell_\LB(a_i)$&  2 &  2 &  2 & 3 &  2 &  1 \\ \hline
 $\ell_\UB(a_i)$&  3 & 4 &  3 & 5 & 4 &  4 \\\hline
 $\bl_\LB(a_i)$&  0 &  0 &   0 & 1 &  1 &   0 \\ \hline
 $\bl_\UB(a_i)$&  1 & 1 &   0 & 2 & 1 &   0 \\\hline
 $\ch_\LB(a_i)$&  0 &  1 & 0 & 4 &  3 &  0 \\ \hline
 $\ch_\UB(a_i)$&  3 & 3 &  1 & 6 & 5 &  2 \\\hline
\end{tabular} 

\begin{tabular}{ |  c | c c c c c c   c c c c  c c c |  } \hline
                        & $u_1$ &  $u_2$ &   $u_3$ &   $u_4$ &   $u_5$ &   $u_6$ 
                       & $u_7$ &   $u_8$ &   $u_9$ &   $u_{10}$ &   $u_{11}$ 
                       &   $u_{12}$ &   $u_{23}$ \\\hline 
 $\bl_\LB(u_i)$&  0 &  0 &   0 & 0 &  0 &   0
                       & 0 &   0 &  0 &   0 &  0 &  0 &  0 \\ \hline
 $\bl_\UB(u_i)$&  1 & 1 &   1 & 1 & 1 &   0
                       & 0 &   0 &  0 &   0 &  0 &  0 &  0\\\hline
 $\ch_\LB(u_i)$&  0 &  0 &   0 & 0 &  1 &   0
                       & 0 &   0 &  0 &   0 &  0 &  0 &  0 \\ \hline
 $\ch_\UB(u_i)$&  1 & 0 &   0 & 0 & 3 &   0
                       & 1 &   1 &  0 &   1 &  2 & 4 &  1 \\\hline
\end{tabular} 

\begin{tabular}{ |  c | c c c c c c   c c c c c c  c c c c c |  } \hline
                               & $a_1$ &  $a_2$ &   $a_3$ &   $a_4$ &   $a_5$ &   $a_6$ 
                               & $a_7$ &  $a_8$ &   $a_9$ &   $a_{10}$ &   $a_{11}$ &   $a_{12}$ 
                               & $a_{13}$ &   $a_{14}$ &   $a_{15}$ &   $a_{16}$ &   $a_{17}$  \\\hline
 $\bd_{2, \LB}(a_i)$ &  0    &  0 &   0 & 1 &  0 &   0
                                &  0   &  0 &  0 & 0 &  0 &   1
                                &  0    &  0 &   0 & 0     & 0  \\ \hline
 $ \bd_{2, \UB}(a_i)$&  1    & 1 &   0 & 2  & 2 &   0  
                                &  0    & 0&   0 & 0 &  0 &   1
                                &  0    &  0 &   0 & 0   & 0   \\ \hline
 $\bd_{3, \LB}(a_i)$ &  0    &  0 &   0 & 0 &  0 &   0
                                &  0   &  0 &  0 & 0 &  0 &   0
                                &  0    &  0 &   0 & 0   & 0   \\ \hline
 $ \bd_{3, \UB}(a_i)$&  0    & 0 &   0 & 0  & 1 &   0 
                                &  0    &  0 &   0 & 0 &  0 &   0
                                &  0    &  0 &   0 &  0    & 0   \\ \hline
\end{tabular} 
\label{table:interior-spec}  
\end{table}

\begin{figure}[h!] \begin{center}
\includegraphics[width=.58\columnwidth]{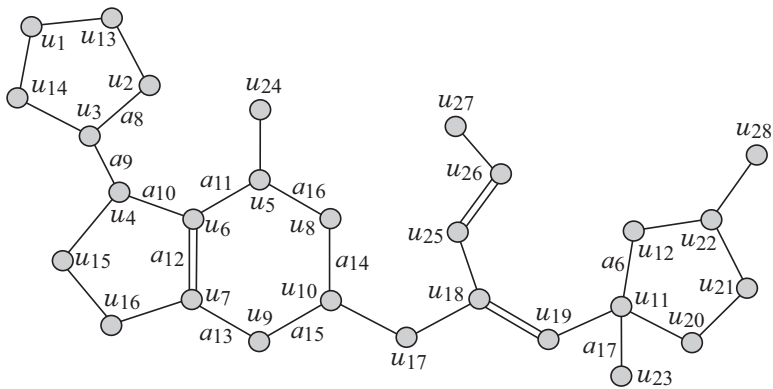}
\end{center} \caption{
An illustration of a graph 
$H^*$ that is obtained from  the seed graph  $\GC$ in 
Figure~\ref{fig:specification_example_1}
under the interior-specification  $\sint$ in 
Table~\ref{table:interior-spec}.    }
\label{fig:specification_example_3} \end{figure}

Figure~\ref{fig:specification_example_3} illustrates an example of 
an $\sint$-extension $H^*$ of seed graph  $\GC$ in 
Figure~\ref{fig:specification_example_1}
under the interior-specification  $\sint$ in 
Table~\ref{table:interior-spec}.  

\subsection*{Chemical-specification}
 
 Let $H^*$ be a graph that serves as 
 the interior $H^\inte$ of a target chemical graph $\C$,
 where the bond-multiplicity of each edge in $H^*$ has be determined.
 Finally we introduce a set of rules for constructing 
   a target chemical graph $\C$ from $H^*$ 
   by choosing  a chemical element $\ta\in \Lambda$ 
and assigning a ${\rho}$-fringe-tree $\psi$
 to each interior-vertex $v\in V^\inte$. 
We introduce the following rules for specifying
the size of $\C$, a set of chemical rooted trees  
that are allowed to use as  ${\rho}$-fringe-trees 
and lower and upper bounds on the frequency of
a chemical element, a chemical symbol, 
and an edge-configuration,
where we call the set of prescribed constants   
 a  {\em chemical specification} $\sce$:   
 
\begin{enumerate}[leftmargin=*]
\item[-] 
Lower and upper bounds $n_\LB,  n^*\in \mathbb{Z}_+$
on the number of vertices, where $\nint_\LB \leq n_\LB\leq n^*$.
 
\item[-] 
Subsets  $\mathcal{F}(v) \subseteq \mathcal{F}(D_\pi), v\in \VC$ 
and $\mathcal{F}_E \subseteq \mathcal{F}(D_\pi)$ 
 of chemical rooted trees $\psi$ with $\h(\anpsi)\leq {\rho}$, where 
 we require that 
 every ${\rho}$-fringe-tree $\C[v]$ rooted at a vertex $v\in \VC$
 (resp., at an internal vertex $v$ not in $\VC$)   in  $\C$ 
 belongs to $\mathcal{F}(v)$ (resp.,   $\mathcal{F}_E$).  
Let  $\mathcal{F}^*:=\mathcal{F}_E\cup \bigcup_{v\in \VC}\mathcal{F}(v)$
and 
$\Lambda^\ex$ denote the set of  chemical elements assigned to non-root
vertices over all chemical rooted trees in $\mathcal{F}^*$.  
 
\item[-] 
A subset  $\Lambda^\inte\subseteq \Lambda^\inte(D_\pi)$, where 
 we require that every chemical element $\alpha(v)$ 
 assigned to an interior-vertex  $v$ in $\C$ belongs to $\Lambda^\inte$.
Let $\Lambda:= \Lambda^\inte\cup \Lambda^\ex$ and
 $\na_\ta(\C)$ (resp., $\na_\ta^\inte(\C)$ and $\na_\ta^\ex(\C)$) 
 denote the number of vertices   (resp.,   interior-vertices and  exterior-vertices)
  $v$ such that $\alpha(v)=\ta$   in  $\C$.
 
\item[-] 
A set $\Ldg^\inte\subseteq \Lambda\times [1,4]$  of chemical  symbols
and  a set $\Gamma^\inte \subseteq \Gamma^\inte(D_\pi)$  
of  edge-configurations  $(\mu,\mu' ,m)$ with $\mu \leq \mu'$, where 
 we require that the edge-configuration $\ec(e)$ of an interior-edge $e$ in $\C$ 
 belongs to $\Gamma^\inte$.
We do not distinguish  $(\mu,\mu' ,m)$ and $(\mu' , \mu,m)$.

\item[-] 
Define  $\Gac^\inte$ to be the set of   adjacency-configurations such that  
$\Gac^\inte:=\{(\ta, \tb, m) \mid (\ta d, \tb d',m)\in \Gamma^\inte\}$.   
Let  $\ac_\nu^\inte(\C), \nu\in \Gac^\inte$   
denote  the number of  interior-edges $e$ such that $\ac(e)=\nu$  in $\C$.
  
\item[-] 
 Subsets $\Lambda^*(v)\subseteq \{\ta\in \Lambda^\inte\mid \val(\ta)\geq 2\}$, 
 $v\in \VC$,  
 we require that every chemical element $\alpha(v)$ 
 assigned to   a vertex $v\in  \VC$
 in the seed graph  belongs to $\Lambda^*(v)$.  

\item[-] Lower and upper bound functions 
$\na_\LB,\na_\UB: \Lambda\to  [1,n^*]$  and 
$\na_\LB^\inte,\na_\UB^\inte: \Lambda^\inte\to  [1,n^*]$ 
on the number of   interior-vertices  $v$ such that  $\alpha(v)=\ta$  in $\C$. 

\item[-] Lower and upper bound functions  
$\ns_\LB^\inte,\ns_\UB^\inte: \Ldg^\inte\to  [1,n^*]$ 
  on the number of   interior-vertices $v$ such that $\cs(v)=\mu$  in $\C$.   

\item[-] Lower and upper bound functions  
$\ac_\LB^\inte,\ac_\UB^\inte: \Gac^\inte \to  \mathbb{Z}_+$ 
 on the number of  interior-edges $e$ such that $\ac(e)=\nu$  in $\C$. 

\item[-] Lower and upper bound functions  
$\ec_\LB^\inte,\ec_\UB^\inte: \Gamma^\inte \to  \mathbb{Z}_+$ 
 on the number of  interior-edges $e$ such that $\ec(e)=\gamma$  in $\C$.  
 
 \item[-] Lower and upper bound functions  
$\fc_\LB,\fc_\UB: \mathcal{F}^*\to  [0,n^*]$ 
  on the number of   interior-vertices $v$ 
  such that $\C[v]$ is r-isomorphic to $\psi\in \mathcal{F}^*$  in $\C$.   
  
 \item[-] Lower and upper bound functions  
$\ac^\lf_\LB,\ac^\lf_\UB: \Gac^\lf \to  [0,n^*]$ 
  on the number of  leaf-edges $uv$ in $\acC$
  with adjacency-configuration $\nu$.  
\end{enumerate}
 
We call a chemical graph $\C$ that satisfies a chemical specification $\sce$
a {\em $(\sint,\sce)$-extension of $\GC$},
and denote by $\mathcal{G}(\GC, \sint,\sce)$ the set of
all $(\sint,\sce)$-extensions of $\GC$. 

Table~\ref{table:chemical_spec}  shows an example of 
a chemical-specification  $\sce$ to the seed graph  $\GC$
 in Figure~\ref{fig:specification_example_1}. 
 

\begin{table}[h!]\caption{Example~2 of a chemical-specification  $\sce$.  
}
\begin{tabular}{ |  l |  } \hline
 $n_\LB=30$,  $n^* =50$. \\\hline
  branch-parameter:   ${\rho}=2$  \\\hline
\end{tabular}

\begin{tabular}{ |  l |  } \hline
 Each of sets $\mathcal{F}(v), v\in \VC$ and
 $\mathcal{F}_E$ is set to be \\
 the set $\mathcal{F}$  of chemical rooted trees $\psi$ with $\h(\anpsi)\leq {\rho}=2$
in Figure~\ref{fig:specification_example_1}(b). \\\hline
\end{tabular}

\begin{tabular}{ |  c | c |   } \hline
  $\Lambda=\{ \ttH,\ttC,\ttN,\ttO, \ttS_{(2)},\ttS_{(6)}, \ttP=\ttP_{(5)}\}$ & 
  $\Ldg^\inte =\{ \ttC2 , \ttC3,  \ttC4, \ttN2, \ttN3, \ttO2,
    \ttS_{(2)}2,  \ttS_{(6)}3, \ttP4   \}$  
\\\hline
\end{tabular}

\begin{tabular}{ |  c | l |  } \hline
  $\Gac^{\inte}$ &
  $ \nu_1 \!=\!(\ttC   , \ttC  , 1) ,   \nu_2 \!=\!(\ttC   , \ttC  , 2) ,   
   \nu_3 \!=\!(\ttC   , \ttN  , 1) ,  \nu_4 \!=\!(\ttC  , \ttO  , 1), 
    \nu_5 \!=\! (\ttC, \ttS_{(2)}, 1),\nu_6 \!=\!(\ttC  , \ttS_{(6)}, 1), 
    \nu_7 \!=\! (\ttC  , \ttP  , 1) $  \\ \hline
\end{tabular}

\begin{tabular}{ |  c | l |  } \hline
  $\Gamma^{\inte}$ &
  $ \gamma_1 \!=\! (\ttC 2 , \ttC 2, 1) ,
   \gamma_2 \!=\!(\ttC 2 , \ttC 3, 1) ,  
   \gamma_3 \!=\!(\ttC 2 , \ttC 3, 2) ,  
   \gamma_4 \!=\!(\ttC 2 , \ttC 4, 1) , 
   \gamma_5 \!=\!(\ttC 3 , \ttC 3, 1) , 
   \gamma_6 \!=\!(\ttC 3 , \ttC 3, 2) , $ \\
   &
  $   
    \gamma_7 \!=\!(\ttC 3 , \ttC 4, 1), 
   \gamma_8 \!=\!(\ttC 2 , \ttN 2, 1) ,  
   \gamma_9 \!=\!(\ttC 3 , \ttN 2, 1) ,  
   \gamma_{10} \!=\!(\ttC 3 , \ttO 2, 1), 
    \gamma_{11} \!=\!(\ttC 2 , \ttC 2, 2),  
    \gamma_{12} \!=\!(\ttC 2 , \ttO 2, 1) ,$ \\
   &
  $  
    \gamma_{13} \!=\!(\ttC 3 , \ttN3, 1), 
    \gamma_{14} \!=\!(\ttC 4, \ttS_{(2)} 2, 2),  
    \gamma_{15} \!=\!(\ttC 2 , \ttS_{(6)}3, 1), 
   \gamma_{16} \!=\!(\ttC 3 , \ttS_{\tiny (6)}3, 1), 
    \gamma_{17} \!=\!(\ttC 2, \ttP4, 2), $ \\
   &
  $  
    \gamma_{18} \!=\!(\ttC 3, \ttP4, 1)  
     $ \\ \hline
\end{tabular}
    
\begin{tabular}{ |  l|  } \hline
$\Lambda^*(u_1)=\Lambda^*(u_8)=\{{\tt C,  N}\}$, 
$\Lambda^*(u_9)=\{{\tt C, O}\}$, 
   $\Lambda^*(u)=\{\ttC\}$, $u\in \VC\setminus\{u_1,u_8,u_9\}$
   \\\hline
\end{tabular}

\begin{tabular}{ |  c | c c c c  c c c |  } \hline
                         & ${\tt H}$  & ${\tt C}$ &   ${\tt N}$ &     ${\tt O}$ 
                         & $\ttS_{(2)}$ & $\ttS_{(6)}$ & $\ttP$  \\\hline
 $\na_\LB(\ta)$ & 40 &  27 &  1 &   1 & 0 & 0 & 0   \\ \hline 
 $\na_\UB(\ta)$ & 65 & 37 & 4 &  8  &   1 &   1 &   1 \\\hline
\end{tabular} 
\begin{tabular}{ |  c | c c c  c c c   |  } \hline
   & $\ttC$ &   $\ttN$ &     $\ttO$  & $\ttS_{(2)}$ & $\ttS_{(6)}$ & $\ttP$  \\\hline
 $\na_\LB^{\inte}(\ta)$ &   9 &  1 &   0  & 0 & 0 & 0      \\ \hline
 $\na_\UB^{\inte}(\ta) $&  23 & 4 & 5 &   1 &   1 &   1  \\\hline
\end{tabular} 

\begin{tabular}{ |  c | c c c c c c  c c c   |  } \hline
    & $\ttC2$ &  $\ttC3$ &   $\ttC4$ & $\ttN2$ &   $\ttN3$ &   $\ttO2$
   & $\ttS_{(2)}2$ & $\ttS_{(6)}3$ & $\ttP4$  \\\hline
 $\ns_\LB^{\inte}(\mu)$ &  3 &  5 &   0 & 0 &  0 &   0 & 0 &  0 &   0    \\ \hline
 $\ns_\UB^{\inte}(\mu) $&  8 & 15 & 2 & 2 & 3 &  5  &   1 &   1 &   1   \\\hline
\end{tabular} 

\begin{tabular}{ |  c | c c c c c c c |  } \hline
         & $\nu_1 $ &   $\nu_2 $ & $\nu_3 $   & $\nu_4 $
         &   $\nu_5 $ & $\nu_6 $   & $\nu_7 $ \\\hline
 $\ac_\LB^{\inte}(\nu)$  &  0  &  0  & 0  & 0   & 0  & 0 & 0     \\ \hline
 $\ac_\UB^{\inte}(\nu)$ &  30 & 10 & 10 & 10 & 1 & 1 & 1 \\\hline
\end{tabular} 

\begin{tabular}{ |  c | c c c c c c c c c c c c c c c c c c |  } \hline
    & $\gamma_1 $ &   $\gamma_2 $ & $\gamma_3 $   & $\gamma_4 $ 
     & $\gamma_5 $
    & $\gamma_6 $ &   $\gamma_7 $ & $\gamma_8 $   & $\gamma_9 $ 
     & $\gamma_{10} $   & $\gamma_{11} $       & $\gamma_{12} $          
     & $\gamma_{13} $   & $\gamma_{14} $       & $\gamma_{15} $          
     & $\gamma_{16} $   & $\gamma_{17} $       & $\gamma_{18} $          
                            \\\hline
 $\ec_\LB^{\inte}(\gamma)$ &  0 &  0 & 0 &  0  & 0 &  0 &  0 & 0 &  0  & 0 & 0 & 0 
    &  0 & 0 &  0  & 0 & 0 & 0  \\ \hline
 $\ec_\UB^{\inte}(\gamma) $& 4 & 15 & 4 &  4  & 10 &  5 & 4 & 4 &  6 & 4 & 4 & 4
 &  2 & 2 &  2  & 2 & 2 & 2  \\\hline
\end{tabular}

\begin{tabular}{ |  c | c   c   |  } \hline 
& $\psi\in\{\psi_i\mid i=1,6,11\}$ 
& $\psi\in \mathcal{F}^*\setminus \{\psi_i\mid i=1,6,11\}$ \\\hline
 $\fc_\LB(\psi)$  &  1 &    0   \\ \hline 
 $\fc_\UB(\psi)$ &  10 &  3\\\hline
\end{tabular}

\begin{tabular}{ |  c | c   c   |  } \hline 
& $\nu\in\{(\ttC,\ttC,1),(\ttC,\ttC,2)  \}$ 
& $\nu\in \Gac^\lf \setminus \{(\ttC,\ttC,1),(\ttC,\ttC,2)  \}$   \\\hline
 $\ac^\lf_\LB(\nu)$  &  0 &    0   \\ \hline 
 $\ac^\lf_\UB(\nu)$ &  10 &  8 \\\hline
\end{tabular} 

\label{table:chemical_spec}
\end{table}

Figure~\ref{fig:example_chemical_graph} 
 illustrates an example $\Co$ of 
a   $(\sint,\sce)$-extension of $\GC$   obtained 
from the  $\sint$-extension $H^*$  
 in Figure~\ref{fig:specification_example_3} 
under the chemical-specification $\sce$ in Table~\ref{table:chemical_spec}.  
Note that $\mathrm{r}(\Co)= \mathrm{r}(H^*)= \mathrm{r}(\GC)-1=4$
 holds since the edge in $\Ez$ is discarded in $H^*$.


\section{Test Instances for Inferring Chemical Graphs}\label{sec:test_instances} 

We prepared the following instances (a)-(d) for conducting experiments
of the second phase of the framework. 
 
 In the second phase of inferring chemical graphs, we  use two properties 
 $\pi\in \{${\sc Bp}, {\sc Dc}$\}$ 
 and define a set $\Lambda(\pi)$ of chemical elements as follows:  
 $\Lambda(${\sc Bp}$)=\Lambda(${\sc Dc}$)=\Lambda_7
    =\{\ttH,\ttC,\ttO, \ttN,\ttS_{(2)},\ttS_{(6)},\ttCl \}$. 
 
\begin{itemize} 
  \item[(a)]  $I_{\mathrm{a}} =(\GC,\sint,\sce)$: The instance
  introduced in Section~\ref{sec:specification} to explain the target specification.
For each property $\pi$, we replace
 $\Lambda=\{ \ttH,\ttC,\ttO, \ttN, \ttS_{(2)},\ttS_{(6)},\ttP_{(5)}\}$
in Table~\ref{table:chemical_spec} 
 with $\Lambda(\pi)
 \cap \{\ttH,\ttC,\ttO, \ttN, \ttS_{(2)},\ttS_{(6)},\ttP_{(5)}\}
 =\{\ttH,\ttC,\ttO, \ttN, \ttS_{(2)},\ttS_{(6)}\}$
 and  remove from the $\sce$
 all chemical symbols,  edge-configurations and fringe-configurations
  that cannot be constructed from the replaced element set 
 (i.e., those containing a chemical element  $ \ttP_{(5)}$). 
 \end{itemize}

\begin{itemize} 
  \item[(b)] $I_\mathrm{b}^i=(\GC^i,\sint^i, \sce^i)$, $i=1,2,3,4$:
 An instance for inferring chemical graphs with rank at most 2.  
In the four instances $I_\mathrm{b}^i$, $i=1,2,3,4$, 
the following specifications in $(\sint,\sce)$ are common. 
\begin{enumerate}
\item[] 
Set  $\Lambda:=\Lambda(\pi)$
 for a given property $\pi\in \{${\sc  Bp,  Dc}$\}$, 
 set $\Ldg^\inte$ to be
the set of all possible symbols in $\Lambda\times[1,4]$  
that appear in the data set $\mathcal{C}_\pi$  
and set $\Gamma^\inte$
to be the set  of 
 all  edge-configurations that appear in the data set $\mathcal{C}_\pi$. 
Set  $\Lambda^*(v):= \Lambda$,  $v\in \VC$. 
 
\item[] 
The lower bounds  
 $\ell_\LB $, $\bl_\LB $, $\ch_\LB $,  
 $\bd_{2,\LB}$,   $\bd_{3,\LB}$,  
 $\na_\LB$,  $\na^\inte_\LB$,  $\ns^\inte_\LB$,  
$\ac^\inte_\LB$, $\ec^\inte_\LB$ and $\ac^\lf_\LB$  are all set to be 0.

\item[] 
Set  upper bounds   
 $\na_\UB(\ta):=n^*, \na\in\{\ttH,\ttC\}$,   
 $\na_\UB(\ta):=5, \na\in\{\ttO,\ttN\}$,
 $\na_\UB(\ta):=2, \na\in\Lambda\setminus \{\ttH,\ttC,\ttO,\ttN\}$. 
The other upper bounds  
 $\ell_\UB $, $\bl_\UB $, $\ch_\UB $,  
 $\bd_{2,\UB}$,   $\bd_{3,\UB}$,  
 $\na^\inte_\UB$,  $\ns^\inte_\UB$,  
$\ac^\inte_\UB$, $\ec^\inte_\UB$ and $\ac^\lf_\UB$ 
are all set to be an upper bound $n^*$  on $n(G^*)$.

\item[] 
We specify $n_\LB$ as a parameter and
set
$n^*:=n_\LB+10$,
  $\nint_\LB:=\lfloor (1/4) n_\LB \rfloor$ and
   $\nint_\LB:=\lfloor (3/4) n_\LB \rfloor$. 

\item[] 
   For each property $\pi$, let $\mathcal{F}(\mathcal{C}_\pi)$ denote
    the set of 2-fringe-trees in the compounds in $\mathcal{C}_\pi$,
   and select a subset $\mathcal{F}_\pi^i\subseteq  \mathcal{F}(\mathcal{C}_\pi)$ with
   $|\mathcal{F}_\pi^i|=45-5i$, $i\in [1,5]$.
   For each instance $I_\mathrm{b}^i$, 
   set $\mathcal{F}_E :=\mathcal{F}(v):=  \mathcal{F}_\pi^i$, $v\in \VC$ and 
$\fc_\LB(\psi):=0, \fc_\UB(\psi):=10, \psi\in  \mathcal{F}_\pi^i$. 
\end{enumerate}
 
  Instance $I_\mathrm{b}^1$ is given   by the rank-1 seed graph $\GC^1$ 
  in Figure~\ref{fig:specification_example_polymer}(i)
  and   Instances $I_\mathrm{b}^i$, $i=2,3,4$ are
   given by  the rank-2 seed graph $\GC^i$, $i=2,3,4$ in 
   Figure~\ref{fig:specification_example_polymer}(ii)-(iv).

\begin{itemize} 
 \item[(i)]  For instance $I_\mathrm{b}^1$, select as a seed graph 
  the monocyclic graph   $\GC^1=(\VC,\EC=\Et\cup \Ew)$
  in Figure~\ref{fig:specification_example_polymer}(i),
  where $\VC=\{u_1,u_2\}$, $\Et=\{a_1\}$ and  $ \Ew=\{a_2\}$. 
We  include a linear constraint 
$\ell(a_1)\leq \ell(a_2)$ 
and $5\leq \ell(a_1)+\ell(a_2) \leq 15$  as part of the side constraint. 
  
 \item[(ii)]
 For instance $I_\mathrm{b}^2$, select as a seed graph 
  the  graph   $\GC^2=(\VC,\EC=\Et\cup \Ew\cup \Eew)$ 
  in Figure~\ref{fig:specification_example_polymer}(ii),
  where
$\VC=\{u_1,u_2,u_3,u_4\}$, 
$\Et=\{a_1,a_2\}$, 
$\Ew=\{a_3\}$  and 
$\Eew=\{a_4,a_5\}$. 
%
We include a linear constraint $\ell(a_1)\leq \ell(a_2)$ 
and $\ell(a_1)+\ell(a_2)+\ell(a_3)\leq 15$. 
    
 \item[(iii)]
 For instance $I_\mathrm{b}^3$, select as a seed graph 
  the  graph   $\GC^3=(\VC,\EC=\Et\cup \Ew\cup \Eew)$ 
  in Figure~\ref{fig:specification_example_polymer}(iii),   where
$\VC=\{u_1,u_2,u_3,u_4\}$, 
$\Et=\{a_1\}$, 
$\Ew=\{a_2, a_3\}$  and 
$\Eew=\{a_4,a_5\}$. 
%
We include   linear constraints 
$\ell(a_1)\leq \ell(a_2)+\ell(a_3)$, $\ell(a_2)\leq \ell(a_3)$
and $\ell(a_1)+\ell(a_2)+\ell(a_3)\leq 15$.  

 \item[(iv)] 
 For instance $I_\mathrm{b}^4$, select as a seed graph 
  the  graph   $\GC^4=(\VC,\EC=\Et\cup \Ew\cup \Eew)$ 
  in Figure~\ref{fig:specification_example_polymer}(iv),   where
$\VC=\{u_1,u_2,u_3,u_4\}$, 
$\Ew=\{a_1, a_2, a_3\}$  and 
$\Eew=\{a_4,a_5\}$. 
We   include   linear constraints 
$\ell(a_2)\leq \ell(a_1)+1$,
$\ell(a_2)\leq \ell(a_3)+1$,  $\ell(a_1)\leq \ell(a_3)$  
and $\ell(a_1)+\ell(a_2)+\ell(a_3)\leq 15$. 
 \end{itemize}
 \end{itemize}

 We define instances in (c) and (d) 
 in order to find chemical graphs that have an intermediate structure of
 given two chemical cyclic graphs 
 $G_A=(H_A=(V_A,E_A),\alpha_A,\beta_A)$ 
and $G_B=(H_B=(V_B,E_B),\alpha_B,\beta_B)$.
Let
 $\Lambda_A^\inte$ and  $\Lambda_{\mathrm{dg},A}^\inte$ 
 denote the sets  of chemical elements
 and chemical symbols  of
 the interior-vertices in $G_A$, 
 $\Gamma_A^\inte$   denote the sets of edge-configurations of
  the interior-edges in $G_A$,   
  and 
  $\mathcal{F}_A$ denote the set of 2-fringe-trees in $G_A$.  
Analogously define sets
 $\Lambda_B^\inte$,    $\Lambda_{\mathrm{dg},B}^\inte$,   
 $\Gamma_B^\inte$ and   $\mathcal{F}_B$ 
 in $G_B$.

\begin{itemize}  
\item[(c)]  $I_{\mathrm{c}}=(\GC,\sint,\sce)$: 
An instance aimed to infer a chemical graph $G^\dagger$ such that
the core of $G^\dagger$ is equal to the core of $G_A$ and 
the frequency of each edge-configuration in the non-core of $G^\dagger$
is equal to that of  $G_B$. 
We use chemical compounds CID~24822711 and CID~59170444 in 
 Figure~\ref{fig:instance_I_c_I_d}(a) and (b)
 for $G_A$ and $G_B$, respectively.  \\
Set   a seed graph $\GC=(\VC,\EC=\Eew)$ to be the core of $G_A$. \\
Set  $\Lambda:=\{{\tt H,C,N,O}\}$, 
and  set $\Ldg^\inte$ to be
the set of all possible chemical symbols in $\Lambda\times[1,4]$.\\
Set 
$\Gamma^\inte:=\Gamma_A^\inte\cup \Gamma_B^\inte$ and 
  $\Lambda^*(v):=\{\alpha_A(v)\}$, $v\in \VC$.  \\
Set 
$\nint_\LB:=\min\{\nint(G_A), \nint(G_B)\}$, 
$\nint_\UB:=\max\{\nint(G_A), \nint(G_B)\}$, \\
$n_\LB:=\min\{n(G_A), n(G_B)\}-10=40$ 
and   $n^*:=\max\{n(G_A), n(G_B)\}+5$. \\
Set  lower bounds  
 $\ell_\LB $, $\bl_\LB $, $\ch_\LB $,  
 $\bd_{2,\LB}$,   $\bd_{3,\LB}$,  
 $\na_\LB$,  $\na^\inte_\LB$,  $\ns^\inte_\LB$, 
$\ac^\inte_\LB$ and  $\ac^\lf_\LB$  to be 0.\\
Set  upper bounds   
 $\na_\UB(\ta):=n^*, \na\in\{\ttH,\ttC\}$,   
 $\na_\UB(\ta):=5, \na\in\{\ttO,\ttN\}$,
 $\na_\UB(\ta):=2, \na\in\Lambda\setminus \{\ttH,\ttC,\ttO,\ttN\}$ 
and set the other upper bounds
 $\ell_\UB $, $\bl_\UB $, $\ch_\UB $,  
 $\bd_{2,\UB}$,   $\bd_{3,\UB}$,  
$\na^\inte_\UB$,  $\ns^\inte_\UB$, 
$\ac^\inte_\UB$   and  $\ac^\lf_\UB$ to be  $n^*$. \\
Set $\ec_\LB^\inte(\gamma)$ 
to be the number of core-edges  in $G_A$ with $\gamma\in \Gamma^\inte$ and  
 $\ec_\UB^\inte(\gamma)$  
to be the number interior-edges in $G_A$ and  $G_B$ 
with edge-configuration $\gamma$. \\
Let $\mathcal{F}_B^{(p)}, p\in [1,2]$ denote the set of chemical rooted 
trees r-isomorphic $p$-fringe-trees in $G_B$; \\
Set $\mathcal{F}_E :=\mathcal{F}(v):= 
 \mathcal{F}_B^{(1)}\cup \mathcal{F}_B^{(2)}$, $v\in \VC$ and
$\fc_\LB(\psi):=0, \fc_\UB(\psi):=10, \psi\in \mathcal{F}_B^{(1)}\cup \mathcal{F}_B^{(2)}$. 
 
  \item[(d)] $I_{\mathrm{d}}=(\GC^1,\sint, \sce)$:     
An instance aimed to infer a chemical monocyclic graph $G^\dagger$ such that
the frequency vector of  edge-configurations in  $G^\dagger$
is a vector obtained by merging those of $G_A$ and $G_B$.
We use chemical monocyclic compounds CID~10076784 and CID~44340250
in   Figure~\ref{fig:instance_I_c_I_d}(c) and (d) 
 for $G_A$ and $G_B$, respectively.  
Set a seed graph to be   the monocyclic seed graph  
 $\GC^1=(\VC,\EC=\Et\cup \Ew)$ with 
  $\VC=\{u_1,u_2\}$, $\Et=\{a_1\}$ and  $ \Ew=\{a_2\}$ 
  in Figure~\ref{fig:specification_example_polymer}(i). \\
Set  $\Lambda:=\{{\tt H,C,N,O}\}$,  
 $\Ldg^\inte:=\Lambda_{\mathrm{dg},A}^\inte 
                 \cup \Lambda_{\mathrm{dg},B}^\inte$ and 
$\Gamma^\inte:=\Gamma_A^\inte\cup \Gamma_B^\inte$. \\
Set 
$\nint_\LB:=\min\{\nint(G_A), \nint(G_B)\}$, 
$\nint_\UB:=\max\{\nint(G_A), \nint(G_B)\}$, \\
  $n_\LB:=\min\{n(G_A),n(G_B)\}=40$ and  
  $n^*:=\max\{n(G_A),n(G_B)\}$. \\
Set  lower bounds  
 $\ell_\LB $, $\bl_\LB $, $\ch_\LB $,  
 $\bd_{2,\LB}$,   $\bd_{3,\LB}$,  
 $\na_\LB$,  $\na^\inte_\LB$,  $\ns^\inte_\LB$, 
$\ac^\inte_\LB$  and  $\ac^\lf_\LB$ to be 0.\\
Set  upper bounds   
 $\na_\UB(\ta):=n^*, \na\in\{\ttH,\ttC\}$,   
 $\na_\UB(\ta):=5, \na\in\{\ttO,\ttN\}$,
 $\na_\UB(\ta):=2, \na\in\Lambda\setminus \{\ttH,\ttC,\ttO,\ttN\}$ 
and set the other  upper bounds  
 $\ell_\UB $, $\bl_\UB $, $\ch_\UB $,  
 $\bd_{2,\UB}$,   $\bd_{3,\UB}$,  
  $\na^\inte_\UB$,  $\ns^\inte_\UB$,
$\ac^\inte_\UB$ and  $\ac^\lf_\UB$  to be   $n^*$. \\
For each edge-configuration
 $\gamma \in \Gamma^\inte$,  
let  $\x^*_A(\gamma^\inte)$  (resp., $\x^*_B(\gamma^\inte)$)   denote
 the number of interior-edges with $\gamma$ in $G_A$ (resp., $G_B$), 
 $\gamma \in \Gamma^\inte$ and   
set \\
$\x^*_{\min}(\gamma):=\min\{\x^*_A(\gamma), \x^*_B(\gamma)\}$, 
 $\x^*_{\max}(\gamma):=\max\{\x^*_A(\gamma), \x^*_B(\gamma)\}$, \\
$\ec_\LB^\inte(\gamma):=
\lfloor (3/4)\x^*_{\min}(\gamma)+(1/4)\x^*_{\max}(\gamma) \rfloor$
and  \\
$\ec_\UB^\inte(\gamma):=
\lceil (1/4)\x^*_{\min}(\gamma)+(3/4)\x^*_{\max}(\gamma) \rceil$. \\
Set $\mathcal{F}_E :=\mathcal{F}(v):=  \mathcal{F}_A\cup \mathcal{F}_B$, 
$v\in \VC$ and 
$\fc_\LB(\psi):=0, \fc_\UB(\psi):=10, \psi\in \mathcal{F}_A\cup \mathcal{F}_B$. \\
We  include a linear constraint 
$\ell(a_1)\leq \ell(a_2)$ 
and $5\leq \ell(a_1)+\ell(a_2) \leq 15$  as part of the side constraint. 
 \end{itemize}

\end{document}